\documentclass[%
 reprint,
 amsmath,amssymb,
 aps,
prb,
 floats,
]{revtex4-1}

\usepackage{subfigure}
\usepackage{graphicx}
\graphicspath{{./}{./images/}{./images_all/}}
\DeclareGraphicsExtensions{.eps}

\usepackage{dcolumn}
\usepackage{bm}

\begin{document}

\preprint{APS/123-QED}

\title{Switching dynamics of a magnetostrictive single-domain 
nanomagnet subjected to stress}

%
%
%

\author{Kuntal Roy}
\email{royk@vcu.edu}
\author{Supriyo Bandyopadhyay}
\affiliation{Department of Electrical and Computer Engineering, Virginia Commonwealth University, Richmond, VA 23284, USA}
\author{Jayasimha Atulasimha}
\affiliation{Department of Mechanical and Nuclear Engineering, Virginia Commonwealth University, Richmond, VA 23284, USA}

\date{\today}

\begin{abstract}
The temporal evolution of the magnetization vector
 of a single-domain magnetostrictive nanomagnet, subjected 
to in-plane stress, is studied by solving the Landau-Lifshitz-Gilbert equation. The stress 
 is
ramped up linearly in time and the switching delay, which is the time it
takes for the magnetization to flip, is computed as a function of the ramp rate.
For high levels of stress, the delay exhibits 
a non-monotonic dependence on the ramp rate, indicating that there is an
{\it optimum} ramp rate to achieve the shortest delay.
 For constant ramp rate, the delay initially decreases with 
 increasing stress but then saturates 
 showing that the trade-off between the 
delay and the stress (or the energy dissipated in switching) becomes less and less favorable 
with increasing stress. 
All of these features are due to a complex interplay between the in-plane and out-of-plane
dynamics of the magnetization vector induced by stress. 
\end{abstract}

\maketitle

\section{\label{sec:introduction}Introduction}

There is significant interest in studying the magnetization reversal dynamics of
 multiferroic (strain-coupled magnetostrictive/piezoelectric bilayer) nanomagnets subjected to stress.
This  has potential applications in ultralow 
power non-volatile magnetic logic and memory \cite{roy11,dsouza11,RefWorks:154,fasha11,RefWorks:167,RefWorks:328,RefWorks:330} because
switching a multiferroic nanomagnet with stress dissipates far less energy than switching it with a  magnetic field or spin transfer 
torque produced by a current \cite{roy11}. As a result, stress-mediated switching can reduce energy 
dissipation 
 in magnetic reversal to the point
where
 non-volatile memory and logic systems can be run by harvesting energy solely from the environment without needing a power 
source or battery! This opens up unique applications in situations where energy is a premium -- such as
in implanted medical devices, structural health monitoring systems, ``wearable'' electronics, and 
space based applications.
 
\begin{figure}
\includegraphics[width=3in]{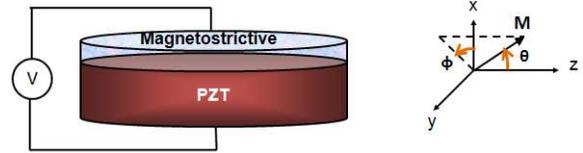}
\caption{\label{fig:multiferroic} An elliptical multiferroic nanomagnet stressed with an applied voltage. 
The polarity of the applied voltage $V$ is determined by the sign of the magnetostrictive coefficient of 
the material used as the magnetostrictive layer. The polarity should be 
such that the stress generated favors aligning the magnetization vector along the in-plane hard axis 
rather than the easy axis.}
\end{figure}

A multiferroic nanomagnet is made of a \emph{magnetostrictive} layer and a  \emph{piezoelectric} 
layer in intimate contact with each other
(see Figure~\ref{fig:multiferroic}). A voltage applied across the piezoelectric layer generates 
in it a mechanical
strain that is mostly transferred to the magnetostrictive layer by elastic 
coupling and produces an extension \cite{RefWorks:164,RefWorks:165} if the 
latter layer is much thinner than the former. If we mechanically constrain the magnetostrictive layer 
from expanding or 
contracting along a certain in-plane direction, e.g. along the minor axis of the ellipse in 
Figure~\ref{fig:multiferroic},
then this will generate uniaxial stress along the major axis through the $d_{31}$ coupling in the piezoelectric.
This stress will cause the magnetization axis of the magnetostrictive layer (nanomagnet) 
to rotate by a 
large angle~\cite{RefWorks:166}, which has been demonstrated in recent experiments~\cite{RefWorks:167}, 
although not in nanoscale.

Let us assume that the shape of the nanomagnet is that of an elliptical cylinder as shown in Figure~\ref{fig:multiferroic}
 and that the initial orientation of the magnetization is close to the major axis
 of the ellipse (z-axis), which is the
magnet's easy axis. In that case, a 90$^{\circ}$ rotation 
will place the magnetization vector along the minor axis, which is the in-plane hard axis. 
Subsequent removal of stress can relax 
it
back to the easy axis, but in a direction 
anti-parallel to the initial direction, resulting in a $\sim$180$^{\circ}$ rotation or ``flip'' 
\cite{roy11}. Ref.~[\onlinecite{roy11}] showed that the energy dissipated
in this process is extremely small ($\sim$200 $kT$ at room temperature) for optimum choice 
of materials, even when the switching takes place in $\sim$1 nanosecond and the stress is turned on
abruptly and instantaneously. In fact, 
a tiny voltage of few mV applied abruptly can generate enough stress to flip
the magnetization \cite{roy11,RefWorks:154} in 1 nanosecond, which results in the miniscule dissipation of 
$\sim$200 $kT$.

In this paper, we are concerned with the following issue. The applied voltage cannot generate strain 
in the magnetostrictive layer 
instantaneously. If we ramp up the voltage gradually with a rise time longer 
than the response time of strain, then 
strain may be able to follow the
voltage quasi-statically. In that case, by controlling the ramp rate of the voltage, we can control 
the rise time of the strain. This may have significant effects on both the switching delay and the 
energy dissipated in the switching process. The purpose of this paper is to investigate this possibility.

Intuitively, one would expect that if the stress is always ramped up to a constant value
regardless of the ramp rate, then the time taken to flip the magnetization (switching
delay), will decrease monotonically with increasing ramp rate. 
The only caveat is that at high stress levels, a very fast ramp rate may cause ripples and ringing in the temporal evolution
of the magnetization vector, which may prolong the switching process. The actual 
situation turns out to be a little more complicated  because the switching dynamics exhibits 
rich and complex behavior 
as a result of the interplay between the in-plane
and out-of-plane excursions of the magnetization vector under application of stress.
This complex interplay has two effects: (1) it makes the switching delay exhibit a non-monotonic dependence 
on the ramp rate when high stresses are encountered, and (2) it makes the switching delay saturate quickly with increasing stress at a 
constant ramp rate.  In this paper, we have studied this intriguing dynamics 
 by solving the Landau-Lifshitz-Gilbert (LLG) equation which governs the temporal evolution of the 
 magnetization vector of a single domain nanomagnet.

The rest of this paper is organized as follows. In Section~\ref{sec:LLG_analytical}, we first derive 
the torque exerted by any applied stress on the magnetization vector of 
the magnetostrictive layer, and then solve the LLG equation analytically in the spherical coordinate 
system to yield 
equations that govern
the time evolution of the magnetization vector. These equations
describe how the polar angle $\theta(t)$
and the 
azimuthal angle $\phi(t)$ of the magnetization vector change with time. 
They
are 
solved numerically to obtain the dynamics of magnetization rotation. In Section~\ref{sec:energy_landscape}, 
we derive the energy landscapes of 
the nanomagnet (energy as a function of magnetization orientation) in the stressed and relaxed
conditions since they are a valuable aid in understanding the time evolution of the 
magnetization vector. In Section~\ref{sec:results}, we present the simulation results,  while in Section~\ref{sec:discussions} we discuss the implications of these results and  present the conclusions.

\section{\label{sec:LLG_analytical}Solution of the Landau-Lifshitz-Gilbert equation}

Consider an isolated nanomagnet in the shape of an elliptical cylinder 
whose elliptical cross section lies in the y-z plane with its major axis aligned along the z-direction and 
minor axis along the y-direction. The dimension of the major axis is $a$, that of the minor axis is
$b$, and the thickness is $l$. The volume of the nanomagnet is $\Omega=(\pi/4)a b l$. 
Let $\theta(t)$ be the angle subtended by the magnetization axis with the +z-axis at any instant of time $t$ and $\phi(t)$ be the angle between the +x-axis and the projection of the magnetization axis on the x-y plane. Thus, $\theta(t)$ is the polar angle and $\phi(t)$ is the azimuthal angle. Note that when $\phi$ = 90$^{\circ}$, the magnetization vector lies in the plane
of the magnet. Any deviation from $\phi$ = 90$^{\circ}$ corresponds to out-of-plane motion.

The total energy of the single-domain nanomagnet (magnetostrictive layer) is the sum of the uniaxial shape anisotropy energy and the 
uniaxial stress anisotropy energy:
\begin{equation}
E(t) = E_{SHA}(t) + E_{STA}(t),
\end{equation}
where $E_{SHA}(t)$ is the uniaxial shape anisotropy energy and $E_{STA}(t)$ is the uniaxial 
stress anisotropy energy at time $t$. The former is given by 
\begin{equation}
E_{SHA}(t) = (\mu_0/2) M_s^2 \Omega N_d(t),
\end{equation}
where $M_s$ is the saturation magnetization and $N_d(t)$ is the demagnetization factor expressed 
as~\cite{RefWorks:157} 
\begin{multline}
N_d(t) = N_{d-zz} cos^2\theta(t) + N_{d-yy} sin^2\theta(t) \, sin^2\phi(t) \\ + N_{d-xx} sin^2\theta(t) \, cos^2\phi(t)
\end{multline}
with $N_{d-zz}$, $N_{d-yy}$, and $N_{d-xx}$ are the components of the demagnetization factor 
along the $z$-axis, $y$-axis, and $x$-axis, respectively. When $a$ and $b$ are nearly equal, 
$l \ll a,b$ and $a > b$, $N_{d-zz}$, $N_{d-yy}$, and $N_{d-xx}$ are approximately given by~\cite{RefWorks:157}
\begin{subequations}
\begin{align}
		N_{d-zz} &= \frac{\pi}{4} \left(\frac{l}{a} \right) 
\left\lbrack 1 - \frac{1}{4}\left(\frac{a-b}{a} \right) - \frac{3}{16}
\left(\frac{a-b}{a} \right)^2 \right\rbrack \\
		N_{d-yy} &= \frac{\pi}{4} \left(\frac{l}{a} \right) 
\left\lbrack 1 + \frac{5}{4}\left(\frac{a-b}{a} \right) + \frac{21}{16}
\left(\frac{a-b}{a} \right)^2 \right\rbrack\\
	  N_{d-xx} &= 1 - (N_{d-yy} + N_{d-zz}).
\end{align}
\end{subequations}
More accurate expressions for these quantities can be found in Ref.~[\onlinecite{RefWorks:402}].

Note that uniaxial shape anisotropy will favor lining up the magnetization along the major axis 
($z$-axis) by minimizing $E_{SHA}$, which is why we will call the major axis the ``easy axis'' 
and the minor axis (y-axis) the ``hard axis'' in the plane of the magnet. 
By mechanically constraining the nanomagnet from expanding or contracting in the y-direction using appropriate clamps,
we will generate uniaxial stress along the $z$-axis (easy axis). 
In that case, the stress anisotropy energy is given by
\begin{equation}
E_{STA}(t) = - (3/2) \lambda_s \sigma(t) \Omega \, cos^2\theta(t),
\label{eq:e_sta}
\end{equation}
where $(3/2) \lambda_s$ is the magnetostriction coefficient of the nanomagnet and $\sigma(t)$ is the stress generated in it by an external agent. Note that a positive $\lambda_s \sigma(t)$ product will favor alignment of the magnetization along the 
major axis ($z$-axis), while a negative $\lambda_s \sigma(t)$ product will favor alignment along the minor 
axis ($y$-axis), because that will minimize $E_{STA}(t)$. In our convention, a compressive stress is negative and tensile stress is positive. Therefore, in a material like Terfenol-D that has positive $\lambda_s$, a compressive stress will favor alignment along the minor axis, and tensile along the major axis. The situation will be opposite with nickel and coblat that have negative $\lambda_s$.
 
At any instant of time, the total energy of the nanomagnet can be expressed as 
\begin{equation}
E(t) = E(\theta(t),\phi(t)) = B(t) sin^2\theta(t) + C(t)
\end{equation}
where 
\begin{subequations}
\begin{align}
B_0(t) &= B_0(\phi(t)) = \cfrac{\mu_0}{2} \, M_s^2 \Omega \left\lbrack N_{d-xx} cos^2\phi(t) \right. \nonumber\\
& \qquad\qquad \left. + N_{d-yy} sin^2\phi(t) - N_{d-zz}\right\rbrack \\
B_{stress}(t) &= (3/2) \lambda_s \sigma(t) \Omega \displaybreak[3]\\
B(t) &= B_0(t) + B_{stress}(t) \\
C(t) &= \cfrac{\mu_0}{2} M_s^2 \Omega N_{d-zz} - (3/2) \lambda_s \sigma(t) \Omega.
\end{align}
\end{subequations}
Note that $B_0(t)$ is always positive for our choice of 
geometry, but $B_{stress}(t)$ can be negative or positive in accordance with the sign of the $\lambda _s \sigma(t)$ product.

The magnetization \textbf{M}(t) of the nanomagnet has a constant magnitude at any given temperature but a variable direction,
so that we can represent it by the vector of unit norm $\mathbf{n_m}(t) =\mathbf{M}(t)/|\mathbf{M}| = \mathbf{\hat{e}_r}$ where $\mathbf{\hat{e}_r}$ is the unit vector in the radial direction in spherical coordinate system represented by ($r$,$\theta$,$\phi$). The other two unit vectors in the spherical coordinate system are denoted by $\mathbf{\hat{e}_\theta}$ and $\mathbf{\hat{e}_\phi}$ for $\theta$ and $\phi$ rotations, respectively. The gradient of potential energy at any particular instant of time $t$ is given by
\begin{equation}
\nabla E(t) = \nabla E(\theta(t),\phi(t)) = \cfrac{\partial E(t)}{\partial \theta(t)} \, \mathbf{\hat{e}_\theta} + \cfrac{1}{sin\theta(t)} \,\cfrac{\partial E(t)}{\partial \phi(t)} \, \mathbf{\hat{e}_\phi} 
\end{equation}
\noindent
where
\begin{eqnarray}
\cfrac{\partial E(t)}{\partial \theta(t)} &=& 2 B(t) sin\theta (t) cos\theta(t) - (3/2) \lambda_s 
\Omega cos^2 \theta(t) \cfrac{\partial \sigma(t)}{\partial \theta(t)} \nonumber\\
&=& 2 B(t) sin\theta (t) cos\theta(t) - \sigma_c (t) \cfrac{\partial \sigma(t)}{\partial 
\theta(t)}\\
\cfrac{\partial E(t)}{\partial \phi(t)} &=& -\frac{\mu_0}{2} \, M_s^2 \Omega (N_{d-xx}-N_{d-yy}) 
sin(2\phi(t)) sin^2\theta (t) \nonumber\\
&& \qquad\quad - (3/2) \lambda_s \Omega cos^2 \theta(t) \cfrac{\partial \sigma(t)}{\partial \phi(t)} 
\nonumber \\ 
&=& - B_{0e}(t) \, sin^2\theta (t) - \sigma_c(t) \cfrac{\partial \sigma(t)}{\partial \phi(t)}
\end{eqnarray}
while 
\begin{eqnarray}
B_{0e}(t)&=&B_{0e}(\phi(t))=\cfrac{\mu_0}{2} \, M_s^2 \Omega (N_{d-xx}-N_{d-yy}) sin(2\phi(t)) \nonumber \\
\sigma_c(t) &=& (3/2) \lambda_s \Omega cos^2 \theta(t). 
\label{definitions}
\end{eqnarray}
The torque acting on the magnetization per unit volume due to shape and stress anisotropy is
\begin{eqnarray}
\mathbf{T_E} (t) &=& - \mathbf{n_m}(t) \times \nabla E(\theta(t),\phi(t)) \nonumber\\
&=& - \mathbf{\hat{e}_r} \times \left\lbrack \left(2 B(t) sin\theta(t) cos\theta(t)- 
\sigma_c(t) \cfrac{\partial \sigma(t)}{\partial \theta(t)}\right) \mathbf{\hat{e}_\theta} \right. 
\nonumber\\
&& \quad\quad \left. - \left(B_{0e}(t)  sin\theta (t) + {{\sigma_c(t)}\over{sin \theta(t)}} 
\cfrac{\partial \sigma(t)}{\partial \phi(t)}\right) \mathbf{\hat{e}_\phi} \right\rbrack  
 \nonumber\\
&=& - \left(2 B(t) sin\theta(t) cos\theta(t)- \sigma_c(t) \cfrac{\partial \sigma(t)}{\partial 
\theta(t)}\right) \mathbf{\hat{e}_\phi}  \nonumber\\
&& \quad  - \left(B_{0e}(t) sin\theta (t) + {{\sigma_c(t)}\over{sin \theta(t)}}  \cfrac{\partial 
\sigma(t)}{\partial \phi(t)}\right) \mathbf{\hat{e}_\theta}								 
\label{eq:stress_torque}
\end{eqnarray}
\noindent

This torque causes the magnetization vector to rotate. The magnetization dynamics under the action of this torque is described by the
 Landau-Lifshitz-Gilbert (LLG) equation as follows.
\begin{equation}
\cfrac{d\mathbf{n_m}(t)}{dt} + \alpha \left(\mathbf{n_m}(t) \times \cfrac{d\mathbf{n_m}(t)}
{dt} \right) = \cfrac{\gamma}{M_V} \mathbf{T_E}(t) 
\label{LLG}
\end{equation}
where $\alpha$ is the dimensionless phenomenological Gilbert damping constant, $\gamma = 2\mu_B \mu_0/\hbar$ is the gyromagnetic ratio for electrons and is equal to $2.21\times 10^5$ (rad.m).(A.s)$^{-1}$, $\mu_B$ is the Bohr magneton, and $M_V=\mu_0 M_s \Omega$. In the spherical coordinate system, 
\begin{equation}
\cfrac{d\mathbf{n_m}(t)}{dt} = \cfrac{d\theta(t)}{dt} \, \mathbf{\hat{e}_\theta} + sin \theta (t)\, \cfrac{d\phi(t)}{dt}
\,\mathbf{\hat{e}_\phi}.
\end{equation}
Accordingly,
\begin{equation}
\alpha \left(\mathbf{n_m}(t) \times \cfrac{d\mathbf{n_m}(t)}{dt} \right) = - \alpha sin 
\theta(t) \, \phi'(t) \,\mathbf{\hat{e}_\theta} +  \alpha \theta '(t) \, \mathbf{\hat{e}_\phi}
\end{equation}
where ()' denotes $d()/dt$. This allows us to write
\begin{multline}
\cfrac{d\mathbf{n_m}(t)}{dt} + \alpha \left(\mathbf{n_m}(t) \times \cfrac{d\mathbf{n_m}(t)}
{dt} \right) \\= (\theta '(t) - \alpha sin \theta (t)\, \phi'(t)) \, \mathbf{\hat{e}_\theta} 
+ 
(sin \theta(t) \, \phi ' (t) + \alpha \theta '(t)) \,\mathbf{\hat{e}_\phi}.
\end{multline}
Equating the $\hat{e}_\theta$ and $\hat{e}_\phi$ components in both sides of Equation
(\ref{LLG}), we get
\begin{multline}
\theta ' (t) - \alpha sin \theta(t) \, \phi'(t)  
=\\  -\frac{\gamma}{M_V} \, \left( B_{0e}(t) sin\theta(t)+ {{\sigma_c(t)}\over{sin \theta(t)}}
\cfrac{\partial \sigma(t)}{\partial \phi(t)} \right)
\end{multline}
\begin{multline}
sin \theta (t) \, \phi '(t) + \alpha \theta '(t) =\\ - \frac{\gamma}{M_V} 
\left( 2 B (t)sin\theta(t) cos\theta(t) - \sigma_c(t) \cfrac{\partial \sigma(t)}{\partial \theta(t)} 
\right). 
\end{multline}
Solving the above equations, we get the following coupled equations for the dynamics of $\theta(t)$ and $\phi(t)$.
\begin{widetext}
\begin{eqnarray}
\left(1+\alpha^2 \right) \cfrac{d\theta(t)}{dt} &=& -\frac{\gamma}{M_V} \left\lbrack 
\left( B_{0e}(t) sin\theta(t)
 + {{\sigma_c(t)}\over{sin \theta(t)}}  \cfrac{\partial \sigma(t)}{\partial \phi(t)}\right) + 
 \alpha \left( 2 B(t) sin\theta (t)cos\theta (t) - \sigma_c(t) 
 \cfrac{\partial \sigma(t)}{\partial \theta(t)}\right) \right\rbrack 
 \label{eq:theta_dynamics} \nonumber \\
&&\\
\left(1+\alpha^2 \right) \cfrac{d \phi(t)}{dt} &=& \frac{\gamma}{M_V} 
\left\lbrack \alpha \left(B_{0e}(t)+ {{\sigma_c(t)}\over{sin^2 \theta(t)}} 
\cfrac{\partial \sigma(t)}{\partial \phi(t)}\right) - \left(2 B(t) cos\theta(t) 
- {{\sigma_c(t)}\over{sin \theta(t)}}  \cfrac{\partial \sigma(t)}{\partial \theta(t)}\right)\right\rbrack.
\label{eq:phi_dynamics}
\end{eqnarray}
\end{widetext}

We should note that the Equations~\eqref{eq:stress_torque},~\eqref{eq:theta_dynamics}, 
and~\eqref{eq:phi_dynamics} are not valid when $\sin\theta=0$ ($\theta=0^\circ$ or $\theta=180^\circ$), 
i.e. when the magnetization direction is {\it exactly} along the easy axis. At these points, 
the torque on the 
magnetization vector  given by Equation (\ref{eq:stress_torque}) diverges.
In order to avoid these points, we will assume that the initial orientation of the 
magnet is $\theta = 179^{\circ}$, and switching is deemed to have been completed
 when $\theta = 1^{\circ}$.
This 1$^{\circ}$ deflection could be caused by thermal fluctuations.
Similar assumptions have been made by other authors \cite{RefWorks:403}.

Equation (\ref{eq:stress_torque}) shows that there is an {\it internal feedback} in the system. Stress induces a torque
which produces the rotation ($\theta(t)$, $\phi(t)$). That rotation generates an additional torque through the 
$\partial \sigma/\partial \theta$ and $\partial \sigma/\partial \phi$ terms. That additional torque affects the response. This feedback
mechanism determines the relation between the rotation and stress, and hence the switching delay as a function of stress.

Note from Equation (\ref{eq:stress_torque}) that the torque has contributions due to the dynamic change in stress 
($\partial \sigma(t)/\partial \theta(t)$, $\partial \sigma(t)/\partial \phi(t)$ terms). These contributions  may aid or 
hinder the rotation of the magnetization vector at different times. This is why the switching delay 
will depend on the ramp rate $\partial \sigma/\partial t$. This dependence turns out to be non-monotonic
because of the complex actions of the magnetization vector.

Note from Equation (\ref{definitions}) 
that the term $B_{0e}(t)$ will be negative when $90^\circ < \phi < 180^\circ$. 
In that case, its contribution to  the time rate of change of $\theta$, i.e. $d \theta(t)/dt$,
will be positive, as we can see from Equation \eqref{eq:theta_dynamics}. In other words,
a negative $B_{0e}(t)$ will tend to {\it increase} $\theta$ with time. Since our initial value of 
$\theta$ is 179$^{\circ}$ and the final value is 1$^{\circ}$, we would prefer that $\theta$ will 
always decrease -- and never increase -- with time in order to complete the 
switching in the shortest time. Therefore, a negative value of $B_{0e}(t)$, or equivalently
$\phi$ lying in the interval [$90^\circ, 180^\circ$], is counterproductive since that makes 
$\theta$ increase with time, causing  the 
magnetization to rotate in the {\it wrong} direction, opposite to the preferred direction. This hinders 
switching and increases the switching delay. Therefore, we will always prefer that $B_{0e}(t)$
remains positive, or equivalently 
 $\phi$ remains in the interval [0$^{\circ}$, 90$^{\circ}$] or [180$^{\circ}$, 270$^{\circ}$].
 
 On the other hand, it is clear from Equation (\ref{eq:phi_dynamics}) that a positive $B_{0e}(t)$
 makes a positive contribution to the rate $d \phi/dt$, which will tend to increase $\phi$ with time and 
 make it exceed 90$^{\circ}$. These two counteracting influences of $B_{0e}(t)$ determine the actual switching
dynamics and the resulting switching 
delay.

Another point to note is that 
when the applied stress is sufficiently high, the stress term $B_{stress}(t)$ dominates  the 
term $B(t)$ in Equations (\ref{eq:theta_dynamics}) and (\ref{eq:phi_dynamics}). The term involving $B(t)$
should remain positivein Equation (\ref{eq:theta_dynamics})
	in order to help $d \theta(t)/dt$ remain negative so that the magnetization vector can rotate in the 
right direction. In order to keep $B_{stress}(t)$ negative, we will have to ensure that the 
product $\lambda_s \sigma$ is negative. For materials with positive magnetostriction (e.g. Terfenol-D),
this requires that the stress is negative, while for materials with negative magnetostriction
(e.g. nickel or cobalt), the stress should be positive. Since tensile stress is positive and 
compressive is negative, Terfenol-D will require compressive stress and nickel or cobalt will require 
tensile stress to initiate switching if the magnetization is initially aligned close to the 
easy axis.

\begin{figure}
\centering
\includegraphics[width=3in]{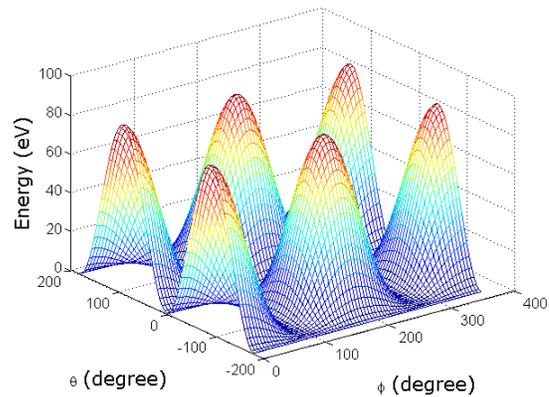}
\caption{\label{fig:energy_landscape_3D} Energy landscape  of a Terfenol-D/PZT multiferroic nanomagnet.
Plot of total energy as a function of polar and azimuthal angles of the magnetization vector.}
\end{figure}

\begin{figure}
\centering
\subfigure[]{\label{fig:energy_landscape_3D_phi_0}\includegraphics[width=3in]{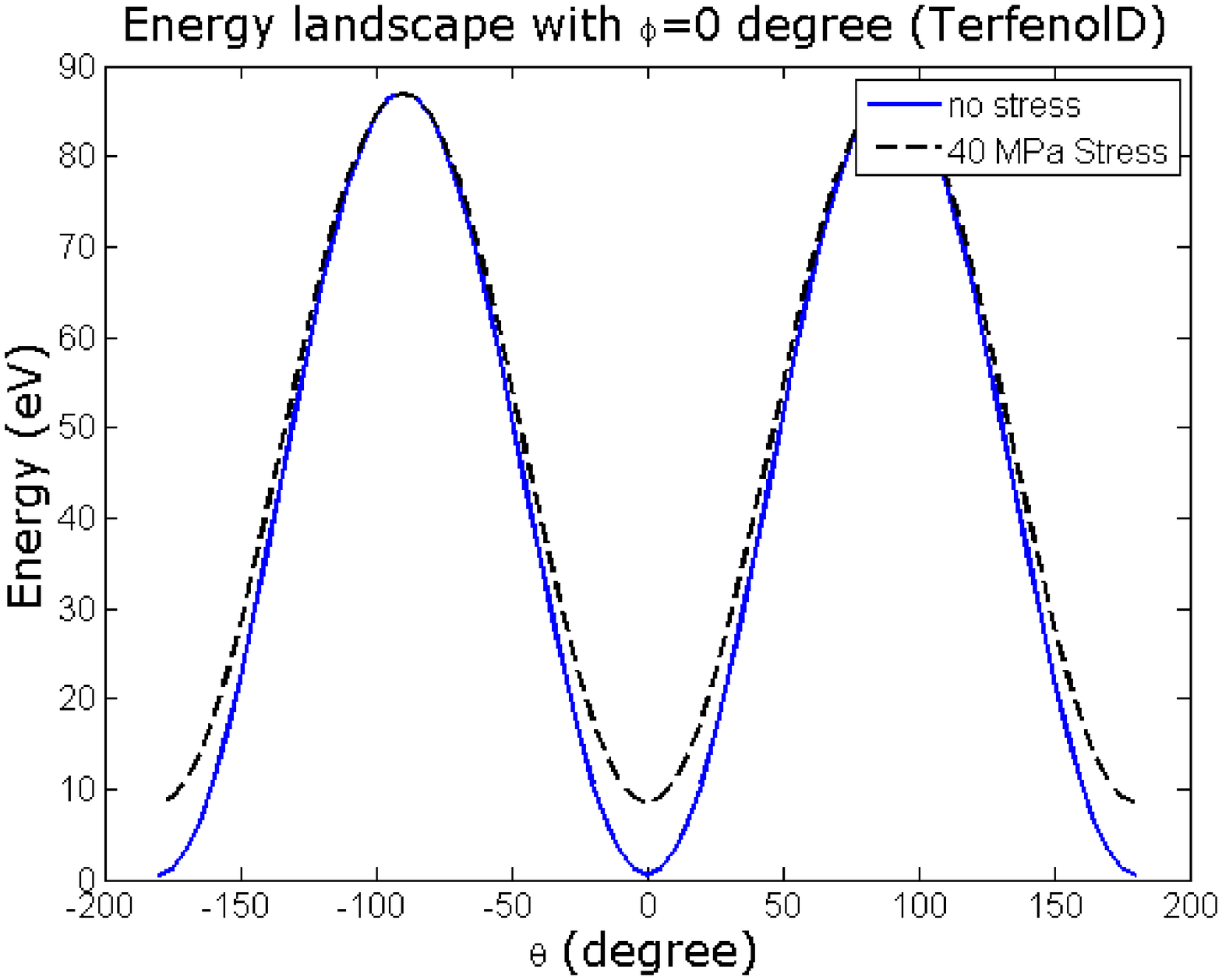}}
\subfigure[]{\label{fig:energy_landscape_3D_phi_90}\includegraphics[width=3in]{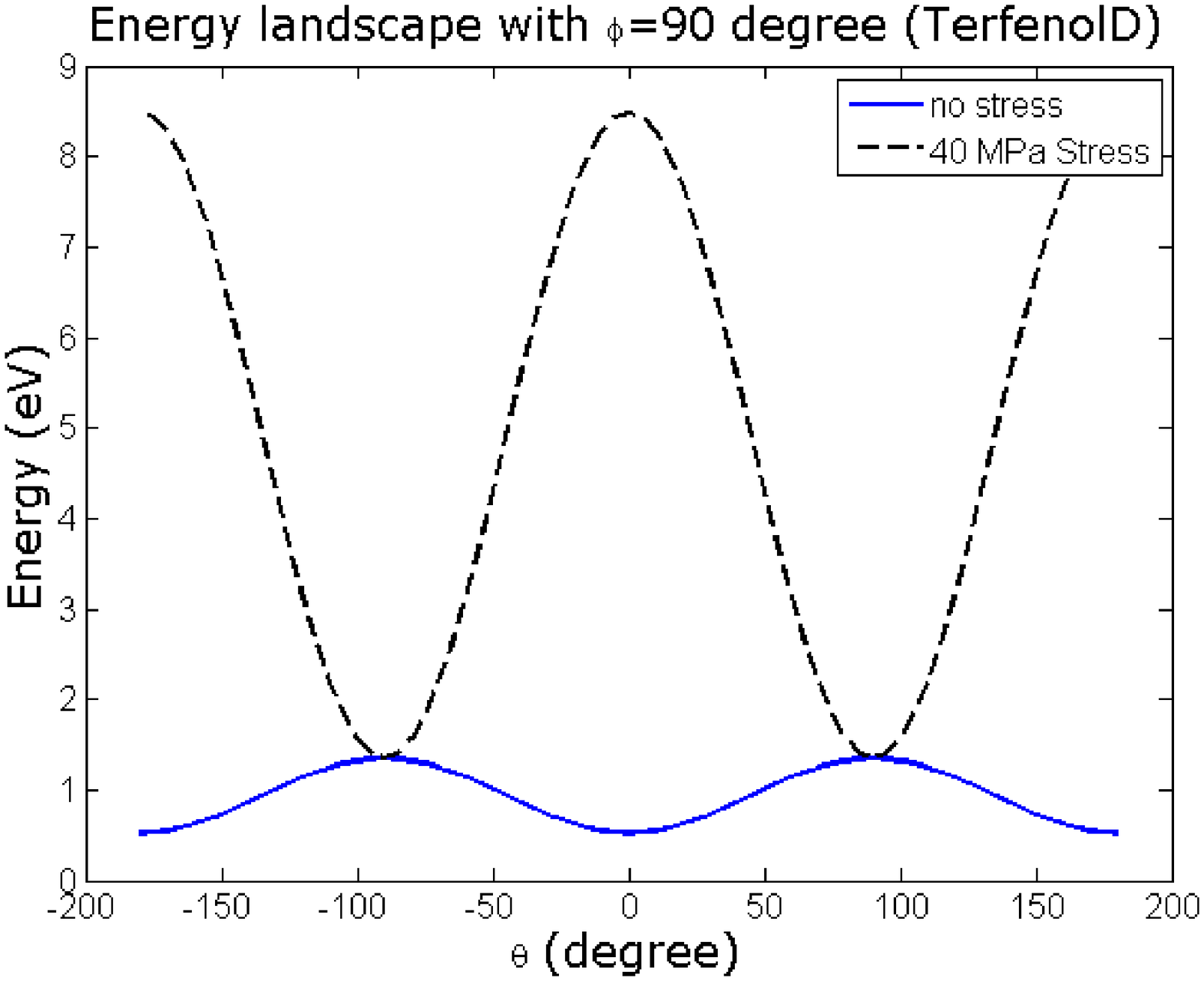}}
\caption{\label{fig:energy_stress_stt} Energy landscape ($E$ versus $\theta$) of a Terfenol-D/PZT multiferroic nanomagnet for 
(a) $\phi=0^\circ$ and (b) $\phi=90^\circ$. The nanomagnet parameters are given in Table I.
The maximum stress that can be generated in a Terfenol-D/PZT multiferroic  nanomagnet 
is of the order of 40 MPa. This stress is enough to depress the shape 
anisotropy barrier and switch the magnetization from the easy axis 
($\theta$ = 0$^{\circ}$, or 180$^{\circ}$) 
to the in-plane hard axis ($\theta$ = 90$^{\circ}$) by shifting the energy minimum 
to $\theta$ = 90$^{\circ}$ when the magnetization vector lies in the plane of the magnet.
The shape anisotropy energy barrier is much taller when $\phi=0^\circ$ than when $\phi$ = 90$^{\circ}$ because of the small thickness of
the nanomagnet making $N_{d-xx}$ much larger than $N_{d-yy}$ or $N_{d-zz}$.}
\end{figure}

\section{\label{sec:energy_landscape}Energy landscape}

The energy landscape of a nanomagnet, which plots the total energy $E(t)$
as a function of the polar angle $\theta(t)$
and the azimuthal angle $\phi(t)$, provides valuable information. The final state of the magnetization 
will 
always be at an energy minimum. Stress will
modify the energy landscape of a nanomagnet and shift the energy minimum from one set of angles 
$ \left ( \phi_i, \theta_i \right )$ to another $ \left ( \phi_f, \theta_f \right )$, thereby 
effecting switching of the magnetization.

Figure~\ref{fig:energy_landscape_3D} shows the energy landscape of a Terfenol-D/PZT 
multiferroic nanomagnet for 
$0^\circ \leq \theta \leq 180^\circ$ and  $0^\circ \leq \phi \leq 360^\circ$ 
without any applied stress. Figures~\ref{fig:energy_landscape_3D_phi_0} 
and~\ref{fig:energy_landscape_3D_phi_90} show how stress modifies the energy landscape
in $\theta$-space for $\phi=0^\circ$ and $\phi=90^\circ$, respectively.

When $\phi=0^\circ$, i.e. the magnetization vector lies in the x-z plane,
the energy barrier separating the two stable magnetizations along the z-axis (easy-axis) 
is $\sim$10 times taller than 
what it is when $\phi=90^\circ$, i.e. when the magnetization vector lies in the y-z plane. 
This happens because  
the thickness of the nanomagnet is much smaller than the other two dimensions, which makes the 
shape anisotropy energy barrier much taller in the former case than in the latter case. 
The stress that can be generated in the magnetostrictive layer by the strained 
piezoelectric layer
is usually sufficient to  rotate the magnetization axis when $\phi=90^\circ$, because the shape anisotropy barrier 
that has to be overcome is small. However, this does not necessarily happen
when $\phi=0^\circ$ because then the shape anisotropy energy barrier is much higher. On the other hand, out-of-plane 
excursion of the magnetization vector generates an additional torque that aids switching. Thus, some out-of-plane excursion 
is beneficial, but if 
the magnetization vector strays out of plane, it encounters a larger energy barrier that prevents switching.
Therefore, the magnetization vector must precess and ultimately return close to the nanomagnet's plane before 
switching can be accomplished. This is the cause of precessional motion.

The energy landscapes allow us to estimate the minimum stress needed to rotate the magnetization vector from the initial orientation
close to
the easy axis to the in-plane hard axis. Once the vector aligns along the in-plane hard axis, stress is removed.
Thereafter,  the magnetization vector will relax back to the easy axis
but to an orientation anti-parallel to the initial orientation. This results in switching. The minimum stress required for this purpose 
is found by
equating the shape anisotropy energy barrier to the stress anisotropy energy, i.e.
\begin{equation}
\left ( N_{d-yy} - N_{d-zz} \right ) M_s^2 {{\mu_0}\over{2}} = {{3}\over{2}} \lambda_s \sigma ,
\end{equation}
which yields
\begin{equation}
\sigma_{min} = \left ( N_{d-yy} - N_{d-zz} \right ) M_s^2{{\mu_0}\over{3 \lambda_s}} .
\end{equation}
However, switching with this minimum stress will incur a very long switching delay, so that some excess
stress will be needed to switch the magnetization reasonably fast. Normally, one would expect the switching 
delay 
to decrease continuously with increasing excess stress, but in reality it saturates beyond a certain stress so that increasing stress
further offers only marginal advantage. This feature cannot be understood from the energy landscape because it is a 
consequence of the interplay between the in-plane and out-of-plane dynamics of the magnetization vector, which is not captured in the
energy profiles.

\begin{table*}
\caption{Material parameters for different materials.}
\begin{center}
\begin{tabular}{c||c|c|c}
& Terfenol-D & Nickel & Cobalt\\
\hline \hline
Major axis (a) & 101.75 nm & 105 nm & 101.75 nm\\
\hline
Minor axis (b) & 98.25 nm & 95 nm & 98.25 nm\\
\hline
Thickness (t) & 10 nm & 10 nm & 10 nm\\
\hline
Young's modulus (Y) & 8$\times$10$^{10}$ Pa & 2.14$\times$10$^{11}$ Pa & 2.09$\times$10$^{11}$ Pa \\
\hline
Magnetostrictive coefficient ($(3/2)\lambda_s$) & +90$\times$10$^{-5}$ & -3$\times$10$^{-5}$  & -3$\times$10$^{-5}$ \\
\hline
Saturation magnetization ($M_s$) & 8$\times$10$^5$ A/m & 4.84$\times$10$^5$ A/m & 8$\times$10$^5$ A/m \\
\hline
Gilbert's damping constant ($\alpha$) & 0.1 & 0.045 & 0.01 \\
\hline\hline
\end{tabular}
\end{center}
\label{tab:material_parameters}
\end{table*}

\section{\label{sec:results}Simulation results}

We consider a multiferroic nanomagnet composed of a PZT layer (lead-zirconate-titanate) and a magnetostrictive layer which is made of either polycrystalline Terfenol-D, or polycrystalline nickel, or polycrystalline cobalt. Because it is polycrystalline, the 
magnetocrystalline layer does not have significant magnetocrystalline anisotropy.
The material parameters for the magnetostrictive layer  are given in 
Table~\ref{tab:material_parameters}~\cite{RefWorks:179,RefWorks:176,RefWorks:178,RefWorks:172, materials}.
They ensure that the shape anisotropy energy barrier is $\sim$32 kT at room temperature. 
The PZT layer is assumed to be four times thicker
than the magnetostrictive layer so that any strain generated in it is transferred almost completely 
to the magnetostrictive layer.
We will assume that the maximum strain that can be generated in the PZT layer is 500 ppm 
\cite{RefWorks:170}, which
would require a voltage of 111 mV because $d_{31}$=1.8e-10 m/V for PZT \cite{pzt2}. 
The corresponding stress is the product of the generated strain ($500\times10^{-6}$) 
and the Young's modulus of the  magnetostrictive layer. Based
on available data for Young's modulus, the maximum allowable stresses for Terfenol-D, nickel, and cobalt 
 are 40 MPa, 107 MPa, and 104.5 MPa, respectively.  

In all our simulations, the initial orientation of the magnetization vector is: $\theta=179^\circ$ and 
$\phi=90^\circ$. Stress is 
applied as a linear ramp and we solve Equations~\eqref{eq:theta_dynamics} and~\eqref{eq:phi_dynamics} 
at each time step. Once  $\theta$ becomes 90$^\circ$, stress is removed and we follow the magnetization 
vector in time until $\theta$ becomes 1$^{\circ}$. 
At that point, switching is deemed to have occurred.

\begin{figure}
\centering
\includegraphics[width=3in]{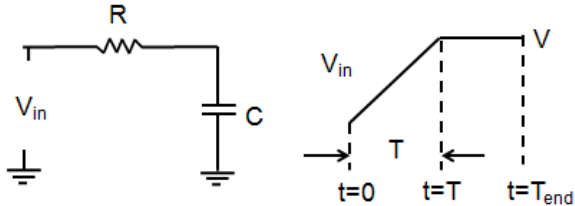}
\caption{\label{fig:RC_circuit} The switching circuit and ramp parameters.}
\end{figure}

We assume that the voltage applied on the piezoelectric is ramped up linearly to its steady-state value 
in time $T$ which we call the {\it rise time}. When the stress is ramped down, we use the 
same rate, i.e. we reduce the stress from its maximum value to zero in time $T$. 
In all cases, the rise time is equal to the fall time.

We also assume that the PZT layer, which acts as a capacitor, is electrically accessed with 
a silver wire of resistivity $\sim$2.6 $\mu\Omega$-cm~\cite{interconnect} so that a typical access 
line of length 10 $\mu m$ and cross section 50 nm $\times$ 50 nm will have a resistance of $\sim$100 $\Omega$. Based on the dimensions of the PZT layer (major axis, minor axis, and thickness), and assuming that the relative dielectric constant 
of PZT is 1000, the capacitance of the PZT layer will be $\sim$2 fF. Therefore, the RC time constant 
associated with charging the capacitor is $\sim$0.2 ps. Since the range of ramp time considered in our 
simulation is 1-150 ps, 
we are in the \emph{adiabatic} limit ($T \gg RC$) and hence the energy dissipation in the external 
circuit that generates the voltage $V$ across the PZT layer
will be less than (1/2)$CV^2$. We assume that the charging circuit is represented by the circuit diagram in 
Figure~\ref{fig:RC_circuit}. The energy dissipated $E_{d,rise}$ during the rise of the voltage 
(charging cycle) for a signal of total time-period $T_{end}$ and ramp-period $T$ can be calculated as below. 
\begin{multline}
E_{d,rise} = CV^2\left(\cfrac{RC}{T}\right) \left\lbrace1-\cfrac{RC}{T} + 
\cfrac{RC}{T} e^{-T/RC} \right. \\ \left.-\cfrac{1}{2}\left(\cfrac{RC}{T}\right)
 \left(1-e^{-T/RC} \right)^2 e^{-2(T_{end}-T)/RC}\right\rbrace, 
\end{multline}
where $C$ is the capacitance of the PZT layer and $V$ is the steady state voltage that generates the 
required stress.
The last term in the above expression comes from a finite value of $T_{end}$. 

The energy dissipated during the discharging cycle
 is $E_{d,fall}$, which can be calculated from an expression similar to the one above, except that the 
 value of $T_{end}$ may be different. For the sake of brevity, we will term the total energy dissipated 
 in the charging circuit $E_{d,rise}+E_{d,fall}$ as the `$CV^2$' energy dissipation in the remainder of 
 this paper.

Because of Gilbert damping in the magnet, an additional energy $E_d$ is dissipated when the nanomagnet switches. 
This energy is given by the expression 
\begin{equation}
E_d = \int_0^{\tau}P_d(t) dt ,
\end{equation}
where $\tau$ is the switching delay and $P_d(t)$, which is the dissipated power, is given by 
\cite{RefWorks:319,RefWorks:124}
\begin{equation}
P_d(t) = \cfrac{\alpha \, \gamma}{(1+\alpha^2) \mu_0 M_s \Omega} \left| T_E(t)\right|^2.
\label{eq:Ed_dissipation}
\end{equation}
	
We sum up the power $P_d(t)$ numerically throughout the switching period to get the corresponding energy 
dissipation $E_d$ and add that to the 
`$CV^2$' dissipation in the switching circuit  to find the total dissipation $E_{total}$. The average power
dissipated during switching is simply $E_d/\tau$.

We analyze the magnetization dynamics of the magnetostrictive layer as a function of both the magnitude 
and the ramp time of the stress for three different materials (Terfenol-D, nickel, cobalt). The results 
are presented in the ensuing subsections. These materials are chosen because of their very different 
material parameters such as Gilbert damping constant, saturation magnetization, and magnetostrictive 
coefficient.

\begin{figure}
\centering
\subfigure[]{\label{fig:theta_dynamics_terfenolD_40MPa_1ps}\includegraphics[width=3in]
{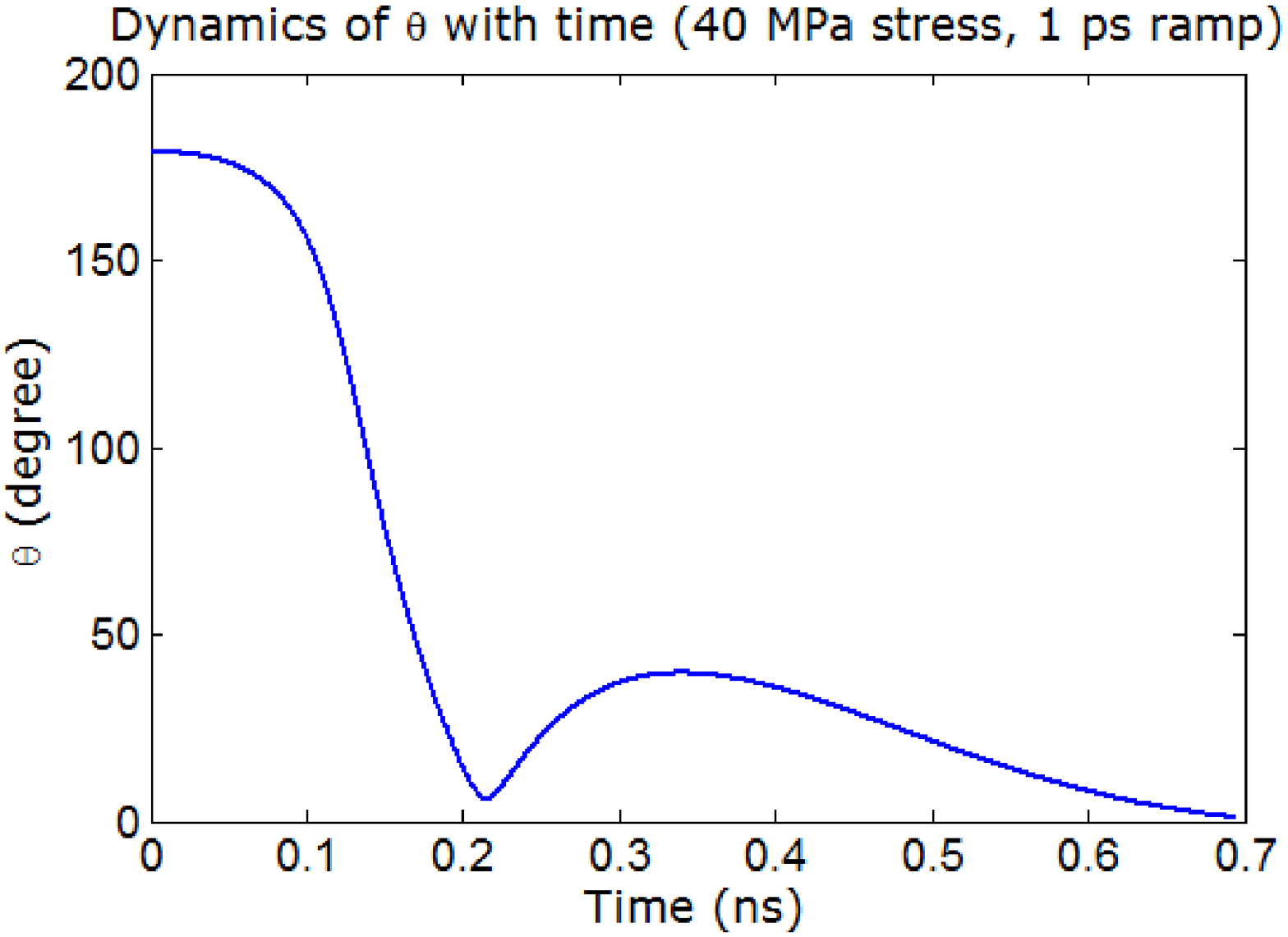}}
\subfigure[]{\label{fig:magnetization_rotation_terfenolD_40MPa_1ps}\includegraphics[width=3in]
{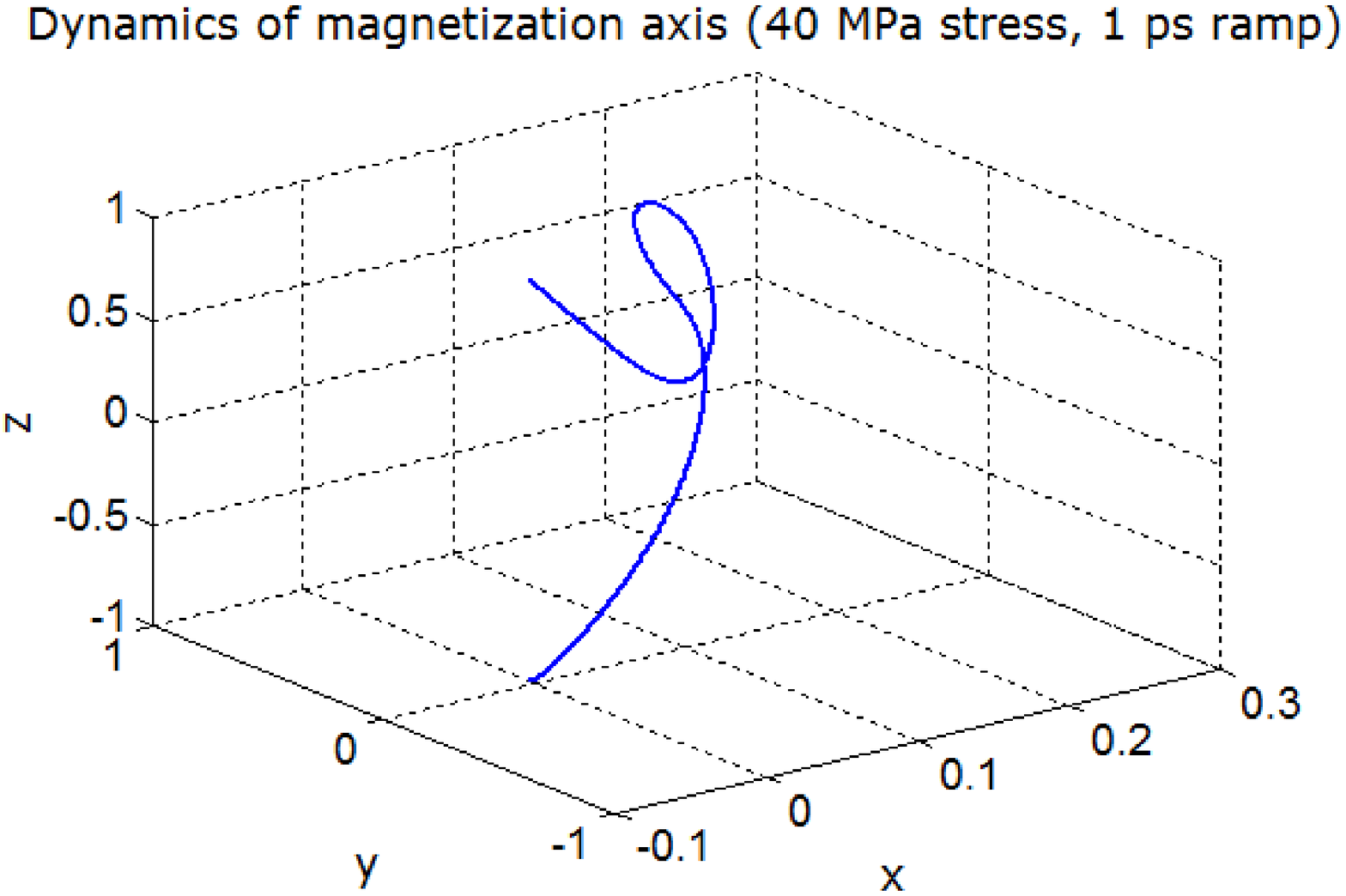}}
\caption{\label{fig:dynamics_terfenolD_40MPa_1ps} Magnetization dynamics in the Terfenol-D/PZT multiferroic 
nanomagnet. The stress is ramped 
up from 0 to 40 MPa in 1 ps: (a) polar angle $\theta$ versus time, and (b) the trajectory traced out by the 
tip of the magnetization vector in three-dimensional space while switching occurs, i.e. during 
the time $\theta$ changes from 179$^{\circ}$ to 1$^{\circ}$. Note that the
magnet's plane is $x$ = 0.}
\end{figure}
\begin{figure}
\centering
\subfigure[]{\label{fig:theta_dynamics_terfenolD_40MPa_60ps}\includegraphics[width=3in]
{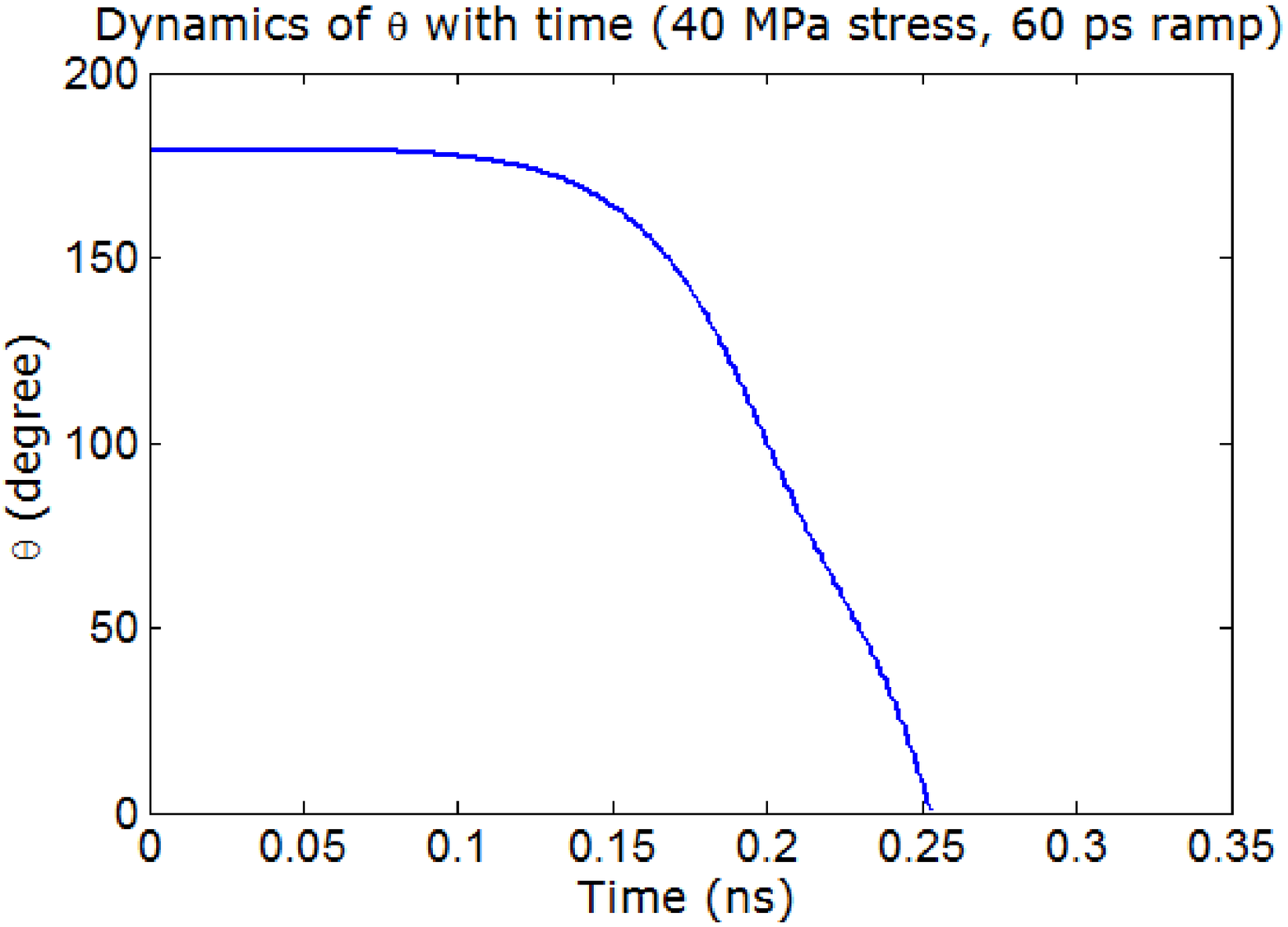}}
\subfigure[]{\label{fig:magnetization_rotation_terfenolD_40MPa_60ps}\includegraphics[width=3in]
{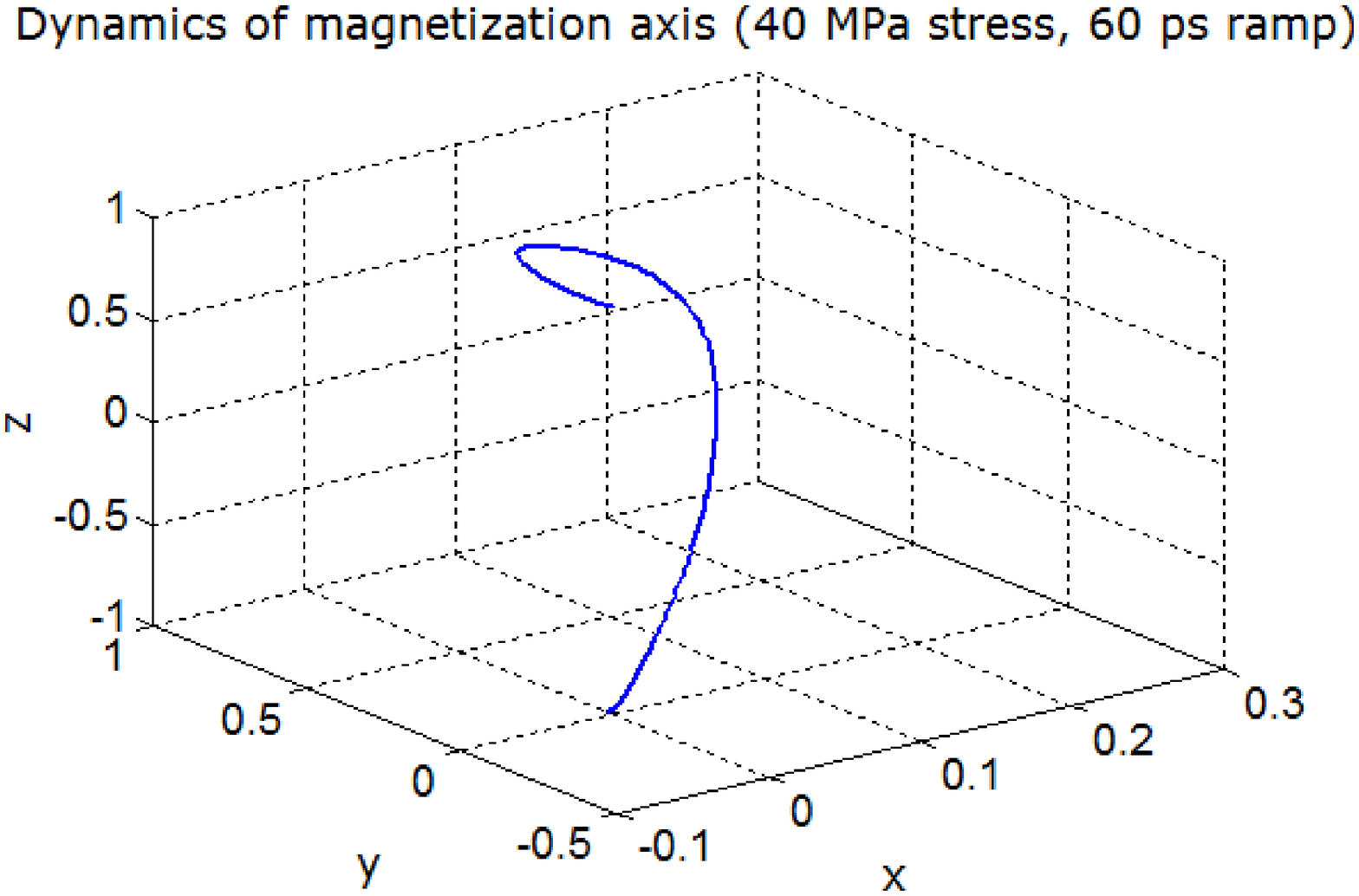}}
\caption{\label{fig:dynamics_terfenolD_40MPa_60ps} Magnetization dynamics in the Terfenol-D/PZT 
multiferroic nanomagnet. The stress is ramped 
up from 0 to 40 MPa in 60 ps: (a) polar angle $\theta$ versus time, and (b) the trajectory traced 
out by the tip of the magnetization vector while switching occurs.}
\end{figure}

\begin{figure}
\centering
\includegraphics[width=3in]{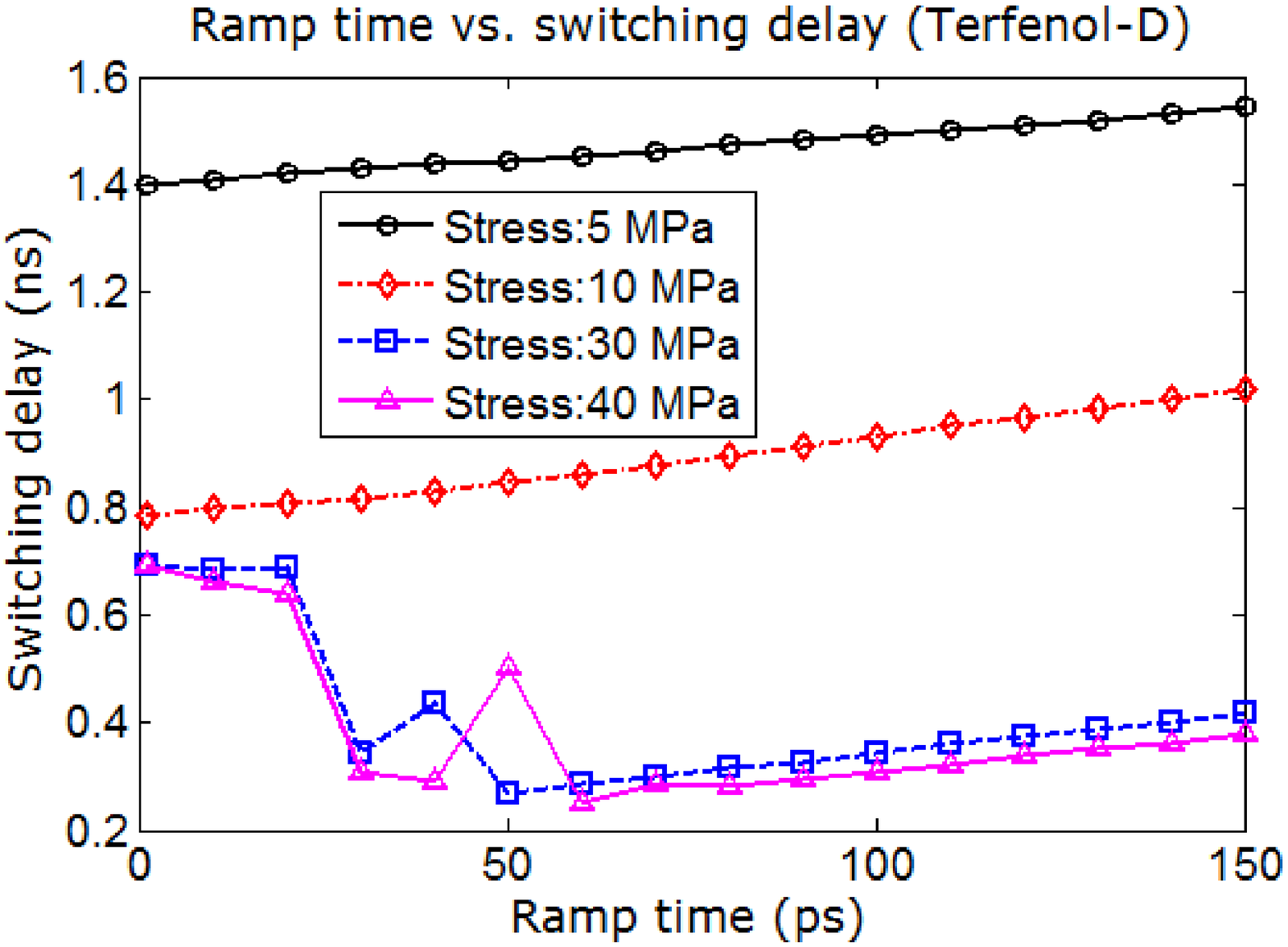}
\caption{\label{fig:delay_ramp_terfenolD} Switching delay of the Terfenol-D/PZT nanomagnet
as a function of the rise (or fall) time of the ramp, 
 with the magnitude of stress as a parameter.}
\end{figure}

\begin{figure}
\centering
\includegraphics[width=3in]{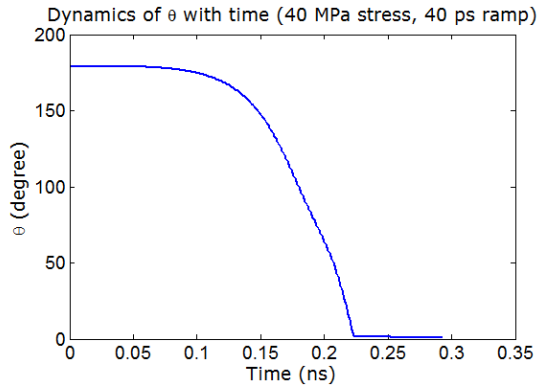}
\caption{\label{fig:theta_dynamics_terfenolD_40MPa_40ps} Magnetization dynamics in the Terfenol-D/PZT 
multiferroic nanomagnet when
the stress is ramped 
up linearly from 0 to 40 MPa in 40 ps. The polar angle $\theta$ is plotted versus time. There is a 
slight ripple, but its amplitude 
is greatly reduced compared to the case when the rise (and fall) time is 1 ps. 
The switching delay in this case is about 285 ps.}
\end{figure}

\begin{figure}
\centering
\includegraphics[width=3in]{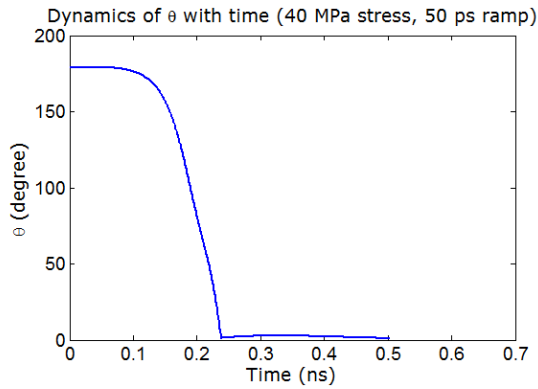}
\caption{\label{fig:theta_dynamics_terfenolD_40MPa_50ps} Magnetization dynamics in the Terfenol-D/PZT 
multiferroic nanomagnet when
the stress is ramped 
up linearly from 0 to 40 MPa in 50 ps. The polar angle $\theta$ is plotted versus time. There is again a 
slight ripple, but its amplitude 
is very small and barely perceptible. The switching delay in this case is about 500 ps.}
\end{figure}

\begin{figure}
\centering
\includegraphics[width=3in]{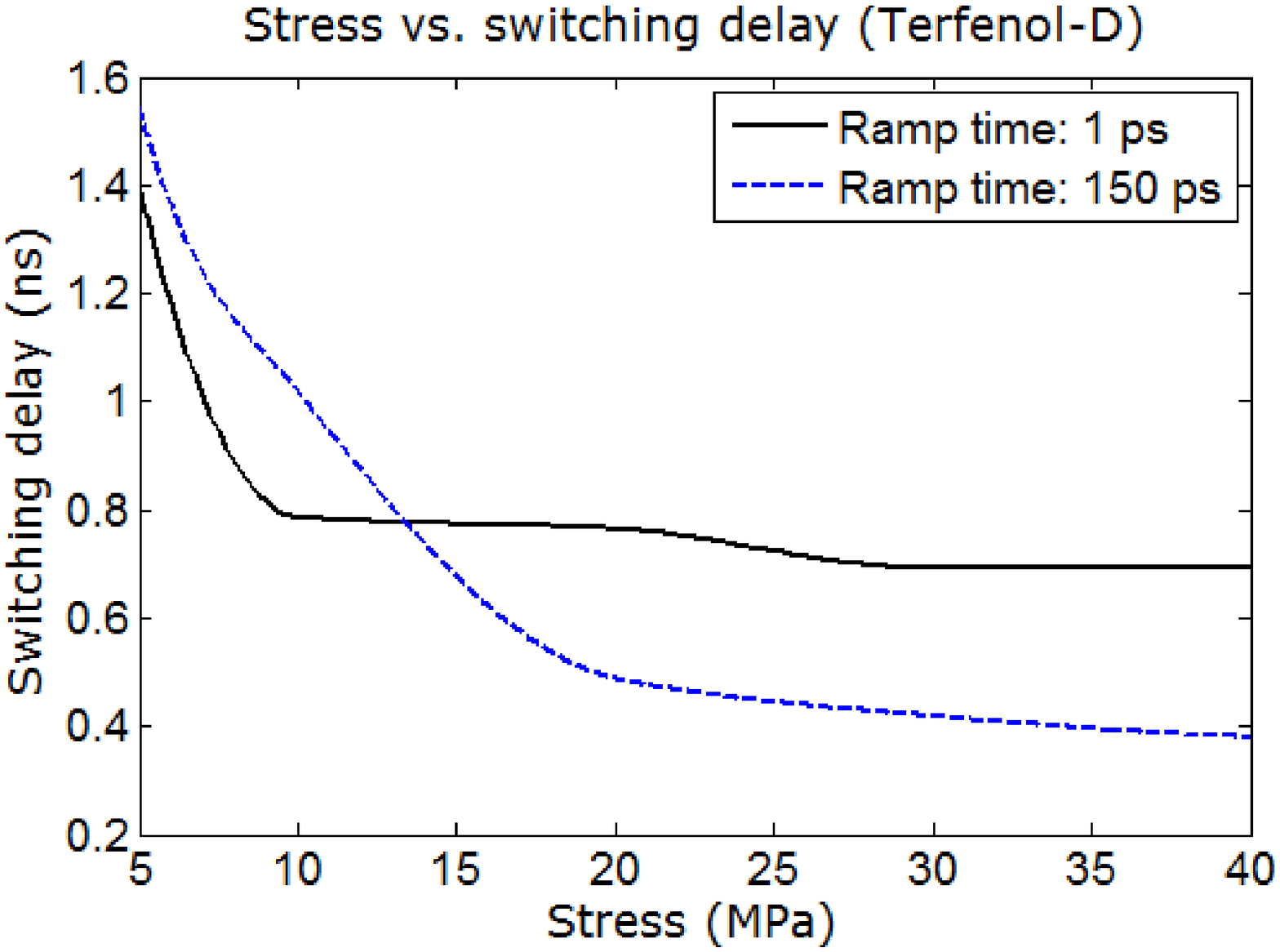}
\caption{\label{fig:delay_stress_terfenolD_ramp_time} Switching delay versus stress for the Terfenol-D/PZT
 multiferroic nanomagnet for two different ramp rise (and fall) times of 1 ps and 150 ps.}
\end{figure}

\begin{figure}
\centering
\includegraphics[width=3in]{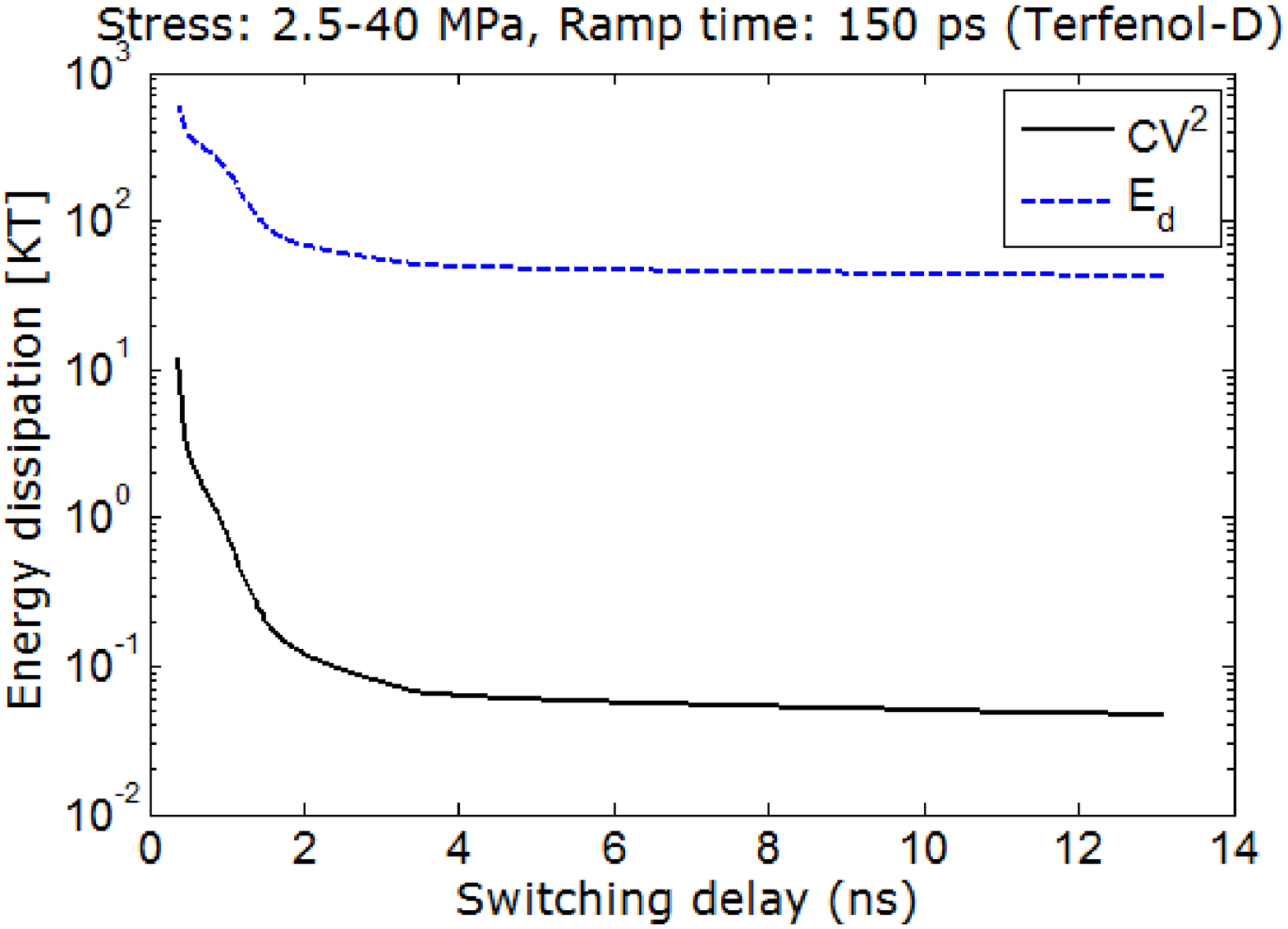}
\caption{\label{fig:delay_energy_terfenolD_stress} Energy dissipated in flipping the magnetization of the Terfenol-D/PZT multiferroic nanomagnet
as a function of switching delay
 for a ramp rise (and fall) time of 150 ps.
This range of switching delay corresponds to a stress range of 2.5 MPa to 40 MPa. The 
energy dissipated in the nanomagnet due to Gilbert damping and the energy dissipated in the external switching 
circuit (`$CV^2$') are shown separately.}
\end{figure}

\begin{figure}[hbtp]
\centering
\includegraphics[width=3in]{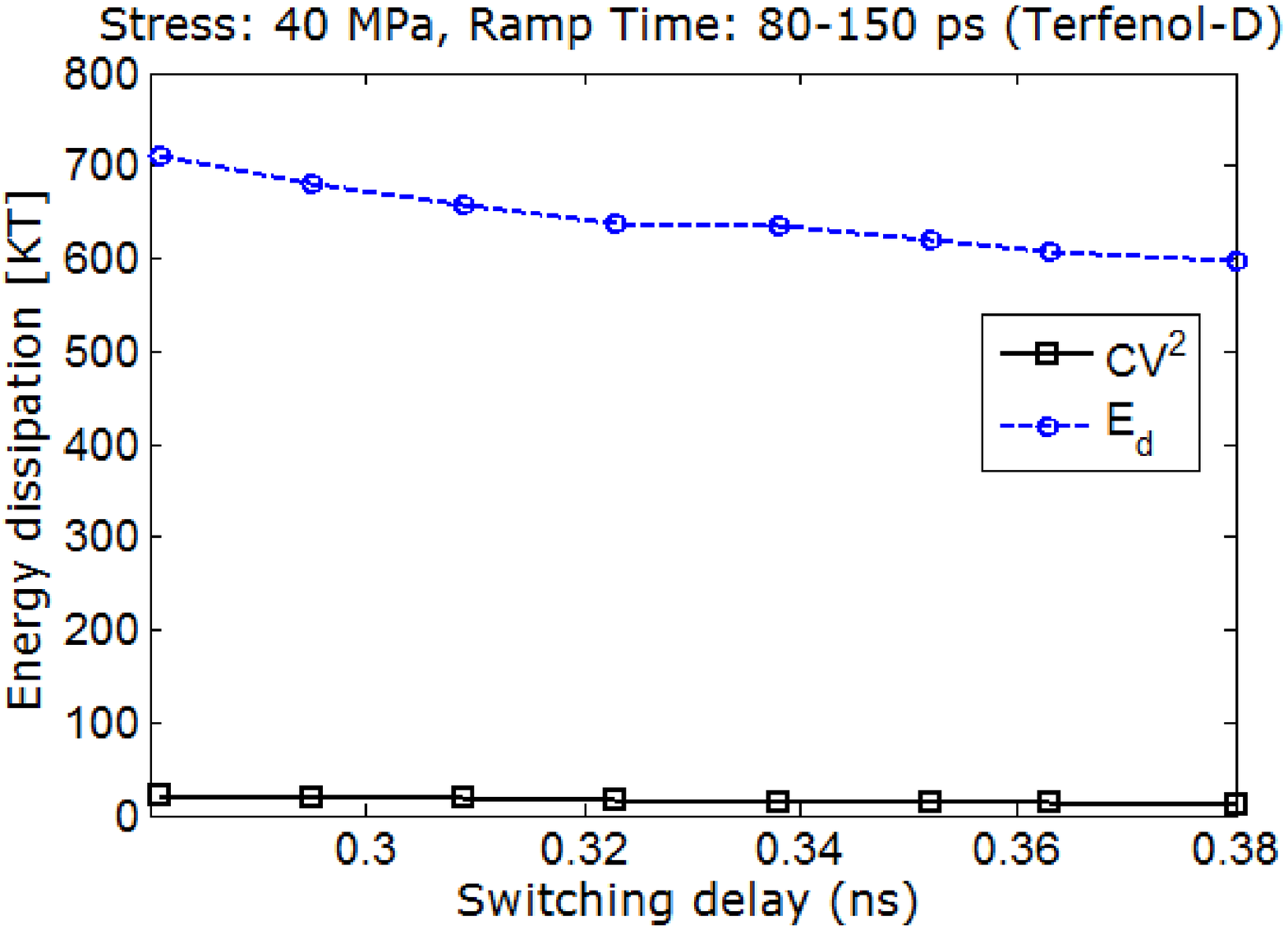}
\caption{\label{fig:delay_energy_terfenolD_ramp_time} For a fixed stress of 40 MPa,
energy dissipated in flipping the magnetization of the Terfenol-D/PZT nanomagnet 
as a function of switching delay when the latter is varied  
by varying the ramp's rise (and fall) time from 80-150 ps.}
\end{figure}

\begin{figure}[hbtp]
\centering
\includegraphics[width=3in]{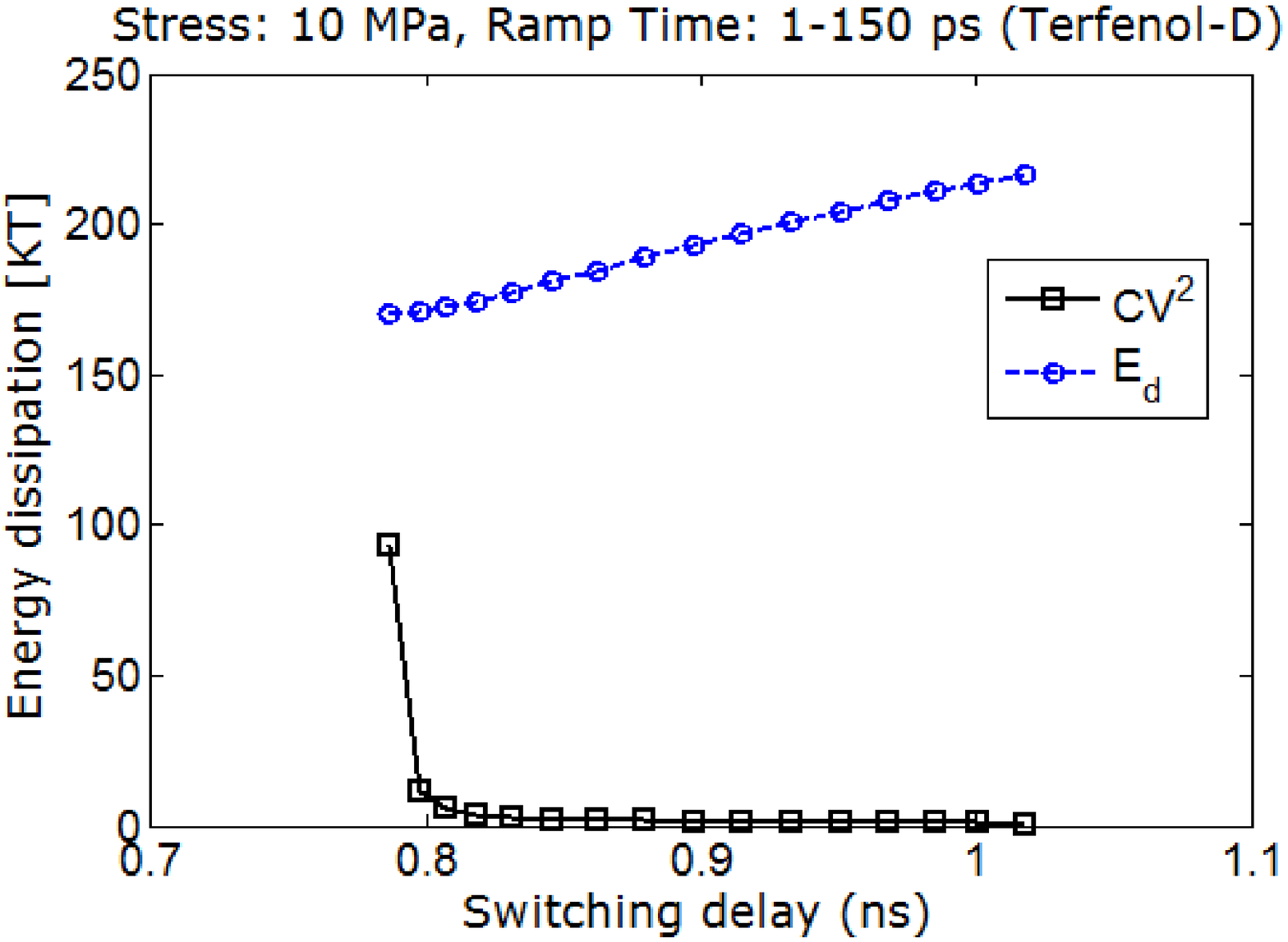}
\caption{\label{fig:delay_energy_terfenolD_ramp_time_10MPa} For a fixed stress of 10 MPa,
energy dissipated in flipping the magnetization of the Terfenol-D/PZT nanomagnet 
as a function of switching delay when the latter is varied  
by varying the ramp's rise (and fall) time from 1-150 ps.}
\end{figure}

\begin{figure}[hbtp]
\centering
\includegraphics[width=3in]{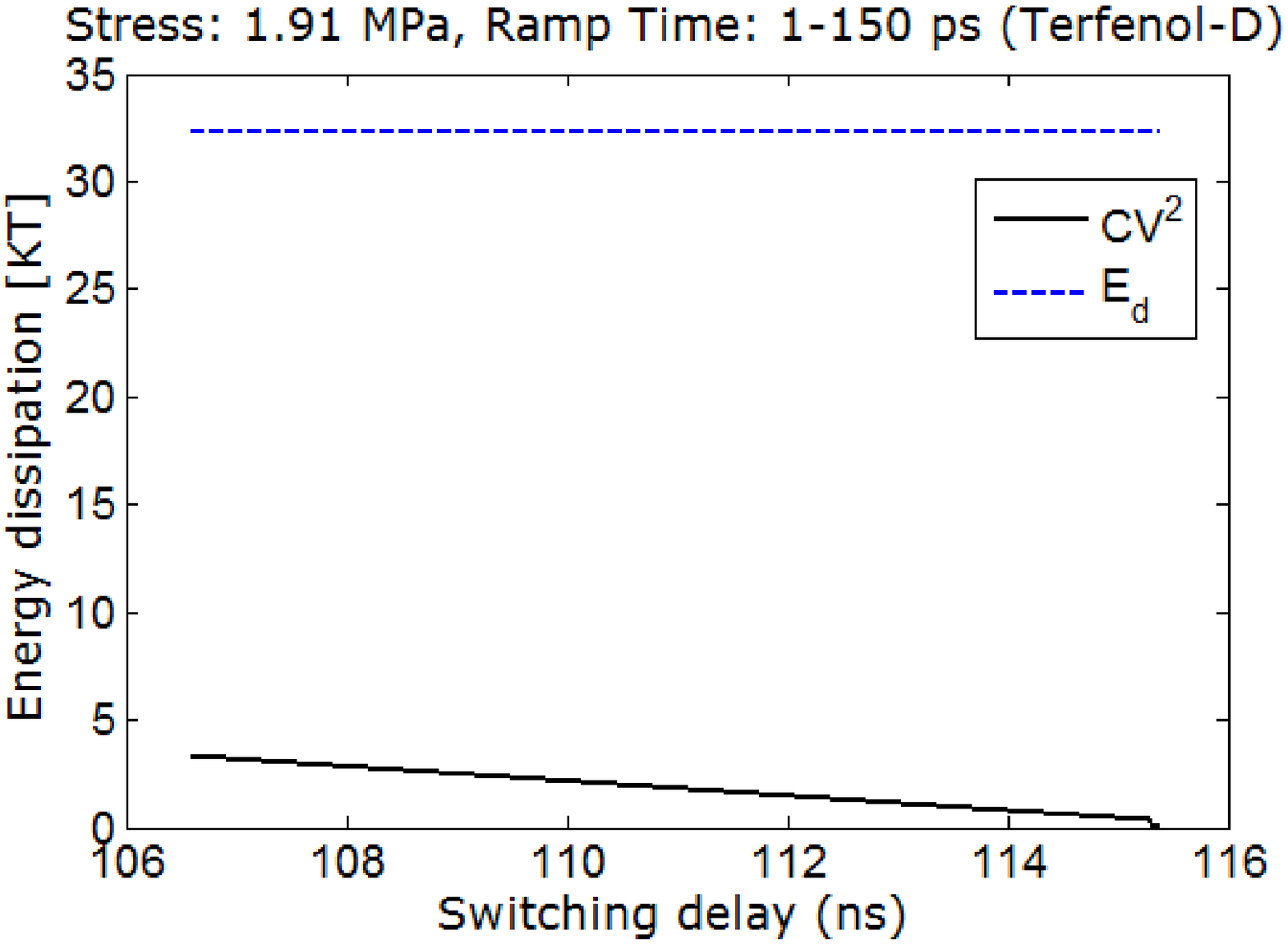}
\caption{\label{fig:delay_energy_terfenolD_ramp_time_1d91MPa} For a fixed stress of 1.91 MPa,
energy dissipated in flipping the magnetization of the Terfenol-D/PZT nanomagnet 
as a function of switching delay when the latter is varied  
by varying the ramp's rise (and fall) time from 1-150 ps.}
\end{figure}

\subsection{Terfenol-D}

Terfenol-D has a positive magnetostrictive coefficient (see Table~\ref{tab:material_parameters}). 
Therefore, we will need a \emph{compressive} stress to rotate the magnetization vector away from
its initial alignment close to easy axis ($\theta$ = 179$^{\circ}$) towards 
$\theta$ = 90$^{\circ}$.  Note that we need to use the correct voltage polarity to ensure that a 
compressive stress is generated on the Terfenol-D layer. The maximum stress that can be generated 
on the Terfenol-D layer with the maximum allowed 500 ppm strain in the PZT layer is 40 MPa, and the 
minimum stress that is needed to switch the nanomagnet is found by equating the stress anisotropy
energy to the shape anisotropy energy barrier. This stress is 1.91 MPa.

\subsubsection{Ramp rate and switching delay}

Equations~\eqref{eq:theta_dynamics} and~\eqref{eq:phi_dynamics}, derived in Section~\ref{sec:LLG_analytical}, 
are solved numerically to find the values of $\theta(t)$ and $\phi(t)$ at any given instant $t$.
This yields the magnetization dynamics under various stresses and ramp rates.

\paragraph{Fast ramp:}

The stress on the Terfenol-D layer is ramped up linearly in time from 0 to the maximum possible value 
of 40 MPa in 1 ps. The corresponding magnetization dynamics is shown in 
Figure~\ref{fig:dynamics_terfenolD_40MPa_1ps}. We notice that the polar  angle $\theta$ continuously 
evolves from its initial value of 179$^{\circ}$ towards its final value of 1$^{\circ}$ for the first 
200 ps. However, because 
of the coupled dynamics of the azimuthal angle $\phi(t)$ and the polar angle $\theta(t)$, $\phi(t)$ 
deviates from its initial value of 90$^{\circ}$ at around 200 ps, forcing the magnetization vector to 
venture out of plane.
 This makes the magnetization vector execute precessional motion in space while its projection on the magnet's plane
changes course and rotates in the direction opposite to the desired 
 direction so that 
$\theta(t)$ begins to increase with time instead of decreasing. Eventually, as the magnetization vector 
stops precessing and returns to
the magnet's plane, which happens around 330 ps, its projection on the magnet's plane starts rotating towards 
its final destination, ultimately reaching 
$\theta$ = 1$^{\circ}$ and $\phi$ = 90$^{\circ}$. Because of the interplay between the $\theta$- and 
$\phi$-dynamics,
which causes the magnetization vector to leave the plane of the nanomagnet after $\sim$200 ps, switching takes around 700 ps.

\paragraph{Slow ramp:}

Figure~\ref{fig:dynamics_terfenolD_40MPa_60ps} shows the magnetization dynamics for a slow ramp that 
takes 60 ps to rise linearly from 0 to 40 MPa. In this case, the magnetization vector does not even budge 
from its initial orientation 
of $\theta$ = 179$^{\circ}$ until the stress reaches its peak value of 40 MPa, which happens after 60 ps 
have elapsed. Thereafter, the magnetization vector rotates towards its final destination of $\theta$ = 
1$^{\circ}$ without ever changing course and rotating in the opposite direction, unlike the previous case.
 The magnetization vector however does leave the magnet's plane in this case as well, but clearly it does 
 not precess as much as in the previous case
(see the trajectory plot). The switching is actually {\it faster} now and takes only 250 ps compared to 
700 ps for the previous case. Thus, a slower ramp can be beneficial  when high stresses are applied. It eliminates the ripple and ringing 
in 
the switching characteristic seen in the previous case by limiting the out-of-plane excursion of the 
magnetization vector and saves precious time.

Figure~\ref{fig:delay_ramp_terfenolD} shows the switching delay as a function of the ramp's rise (or fall) time 
 with the magnitude of stress as a parameter. We see that for stresses of 5 MPa and 10 MPa, the switching delay 
 increases linearly with the ramp time, but for higher stresses of 30 MPa and 40 MPa, the switching delay 
 shows clear non-monotonic behavior for rise (and fall) times less than 60 ps. Normally, the switching delay should 
 increase continuously
 with the ramp's rise (and fall) time,
but  the out-of-plane dynamics 
and precession of the magnetization vector that occurs at very fast ramp rates can reverse this trend and 
cause the 
non-monotonic behavior. Figures~\ref{fig:theta_dynamics_terfenolD_40MPa_40ps} 
and~\ref{fig:theta_dynamics_terfenolD_40MPa_50ps} show the magnetization dynamics for ramp rise (and fall) times 
of 40 ps and 50 ps, respectively. There are some ripples in these cases also, but their 
amplitudes are much reduced so that they are barely visible in the plots. For ramp rise 
times exceeding 60 ps, 
the ripples disappear. Thereafter, the switching delay increases monotonically with the rise (or fall) time.

\subsubsection{Switching delay and energy dissipation}

Figure~\ref{fig:delay_stress_terfenolD_ramp_time} shows the dependence 
 of the switching delay on stress in the Terfenol-D/PZT nanomagnet, with the ramp's rise (or fall) 
 time as a parameter. The two rise (and fall) times considered are 1 ps and 150 ps.
 There is a cross over at 14 MPa stress. At low stress levels below 14 MPa, not much ripple 
 is generated by a fast ramp so that the switching delay is shorter for the faster ramp. At high stress 
 levels exceeding 14 MPa, a fast ramp generates enough ripple that the 
 switching delay becomes longer for the faster ramp. This is the reason for the cross over.
 
Figure~\ref{fig:delay_energy_terfenolD_stress} shows the energy 
dissipated in flipping the magnetization of the Terfenol-D/PZT multiferroic nanomagnet as a function 
of switching delay for a fixed rise (and fall) time of 150 ps. The switching delay is varied by varying the 
stress on the nanomagnet between 2.5 and 40 MPa. The energy dissipated internally in the nanomagnet ($E_d$)
 and the energy dissipated in the switching circuit (`$CV^2$') are shown 
separately. They both tend to saturate at larger delays.

For longer switching delays, the stress needed to flip the magnetization is less and hence the 
voltage $V$ needed to generate the stress is smaller. This leads to a smaller `$CV^2$' dissipation 
in the switching circuit. At the same time, the energy $E_d$ dissipated internally in the nanomagnet is 
smaller when we switch slowly. In this range of switching delay, the energy dissipated in the 
external circuit is much smaller than the energy dissipated internally in the nanomagnet since the 
switching is {\it adiabatic} (the rise (and fall) time is much longer than the RC time constant of the switching 
circuit). The ratios of the two energies however decreases with decreasing switching delay.
Below a switching delay of 4 ns, the energy dissipated internally in the nanomagnet and 
energy dissipated in the switching circuit both increase super-exponentially with decreasing
switching delay. At 1 ns switching delay, the total energy dissipated in switching is only about 200 kT, which
makes this switching methodology {\it extremely energy-efficient}. For this switching delay,
the energy dissipated to switch a state-of-the-art transistor would have been at least two orders of magnitude larger 
\cite{itrs},
and the energy dissipated to switch the same nanomagnet with spin transfer torque will be also at least two orders 
of magnitude larger \cite{RefWorks:329}.

Figure~\ref{fig:delay_energy_terfenolD_ramp_time} plots the energy dissipation as a function of switching
delay for a fixed stress of 
40 MPa. Here, the switching delay is varied by varying the ramp's rise (and fall) time between 80 and 150 ps.
A lower rise (or fall) time is avoided because of the non-monotonic behavior observed in
 Figure~\ref{fig:delay_ramp_terfenolD}. Both $E_d$ and the  `$CV^2$' dissipations
 fall off with increasing switching delay. This means that the average power dissipation falls off
 even more rapidly with increasing switching delay since the energy dissipation is the product of the 
 average power dissipation and the switching delay.  This figure shows that we can switch in
 0.3 - 0.4 ns by dissipating 600 - 700 kT of energy at room temperature.
 
 A similar plot for a lower fixed stress
of  10 MPa is shown in  Figure~\ref{fig:delay_energy_terfenolD_ramp_time_10MPa}. Here, the 
switching delay is varied by varying the rise (and fall) time between 1 ps and 150 ps since there is no issue of 
non-monotonic behavior at such low stress values (no ripples generated in the switching characteristics). 
We see that the energy dissipated internally in the nanomagnet ($E_d$) now increases with increasing
switching delay which is the opposite of the behavior observed in the case of 40 MPa stress. 
In this case, the average power dissipation still goes down with increasing switching delay,
but not fast enough, so that the total energy, which is the product of 
the average power and the switching delay, actually go up with increasing delay. However,
the `$CV^2$' energy dissipation in the external circuit does goes down with increasing delay since switching 
becomes more adiabatic as the delay becomes longer. This figure shows that we can switch in $\sim$1 ns
by dissipating roughly 200 kT of energy.

Figure~\ref{fig:delay_energy_terfenolD_ramp_time_1d91MPa} shows the energy dissipation as 
a function of switching delay for a fixed stress of 1.91 MPa. Once again,
the delay is varied by varying the ramp's rise (and fall) time between 1 and 150 ps. In this case, $E_d$ is nearly
independent of the switching delay, meaning that the average power dissipation in the nanomagnet varies 
inversely with the switching delay. The `$CV^2$' dissipation in the external circuit still goes down
with increasing delay as expected because switching becomes increasingly `adiabatic'. This figure shows that we can switch in 110 ns by dissipating only
$\sim$35 kT of energy at room temperature. The corresponding average power dissipation in this case is 
roughly 35 kT/110 ns = 1.33 pW per nanomagnet per bit flip. If we have an array of magnets with areal
density 10$^{10}$ cm$^{-2}$ (10 Gbits/sq-cm) and 10\% of them are being flipped at any given 
time (10\% activity level), then the power dissipated is 1.3 mW/cm$^2$. The energy needed to run at 
such low power levels can be harvested from the local surroundings using existing energy harvesting 
devices \cite{roundy03,RefWorks:324,RefWorks:326,RefWorks:327} without requiring a battery or external energy source! This opens up
the possibility of unique applications such as medically implanted devices (e.g. processors implanted in a 
patient's brain which  warn of impending epileptic seizures) that run by harvesting energy from a 
patient's body movements without requiring a battery, buoy mounted processors in the open sea that 
harvest energy from swaying motion induced by sea waves, or distributed sensor-processor networks for
structural health monitoring of bridges and buildings that harvest energy from vibrations of the structure 
due to wind or passing traffic. These applications are made possible by the extreme energy efficiency of strain-induced
switching.
\begin{figure}
\centering
\subfigure[]{\label{fig:theta_dynamics_nickel_107MPa_1ps}\includegraphics[width=3in]{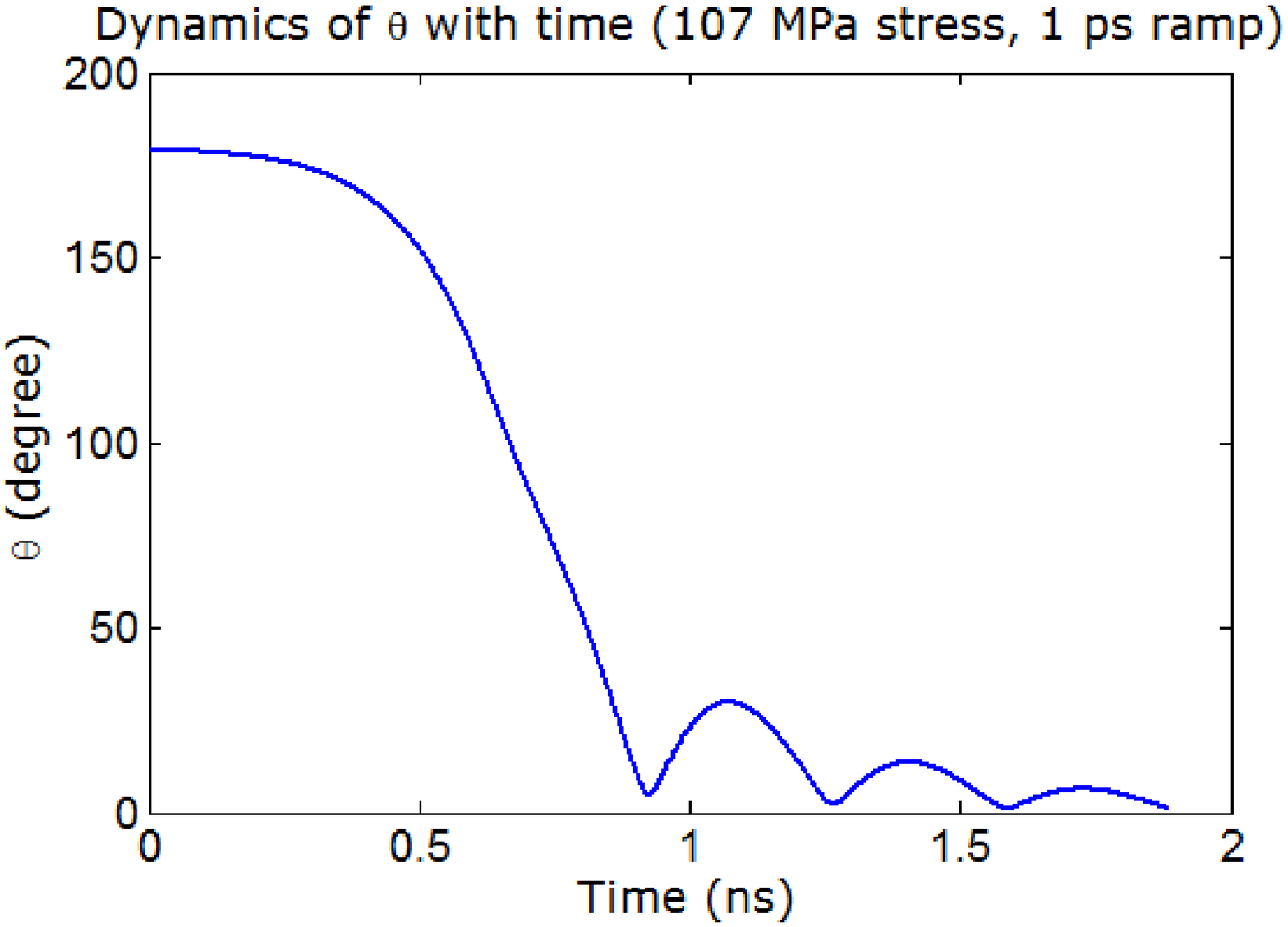}}
\subfigure[]{\label{fig:magnetization_rotation_nickel_107MPa_1ps}\includegraphics[width=3in]{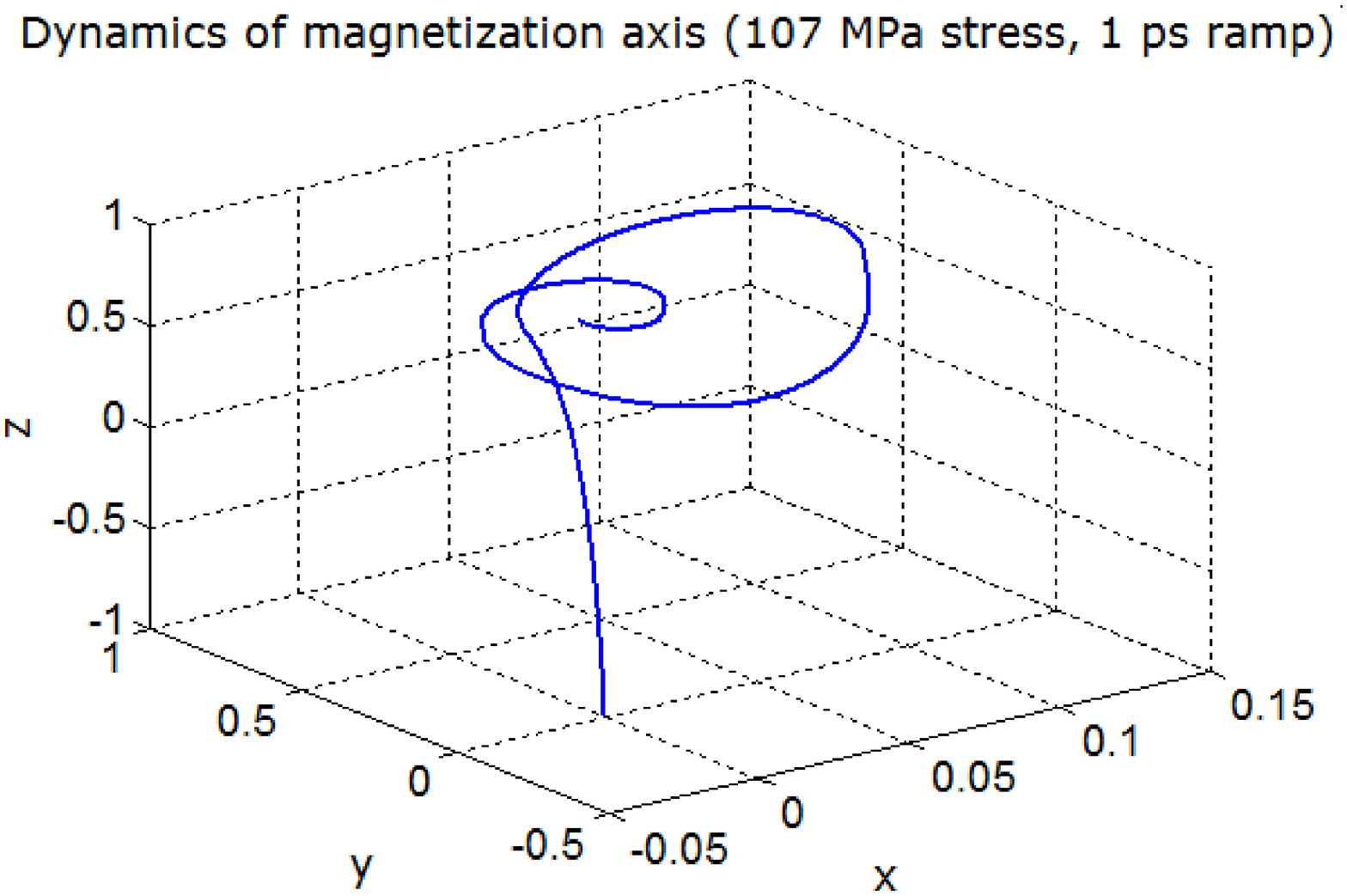}}
\caption{\label{fig:dynamics_nickel_107MPa_1ps} 
Magnetization dynamics in the nickel/PZT multiferroic 
nanomagnet. The stress is ramped 
up from 0 to 107 MPa in 1 ps: (a) polar angle $\theta$ versus time, and (b) the trajectory traced out by the 
tip of the magnetization vector in three-dimensional space while switching occurs, i.e. during 
the time $\theta$ changes from 179$^{\circ}$ to 1$^{\circ}$.}
\end{figure}
\begin{figure}
\centering
\subfigure[]{\label{fig:theta_dynamics_nickel_107MPa_150ps}\includegraphics[width=3in]{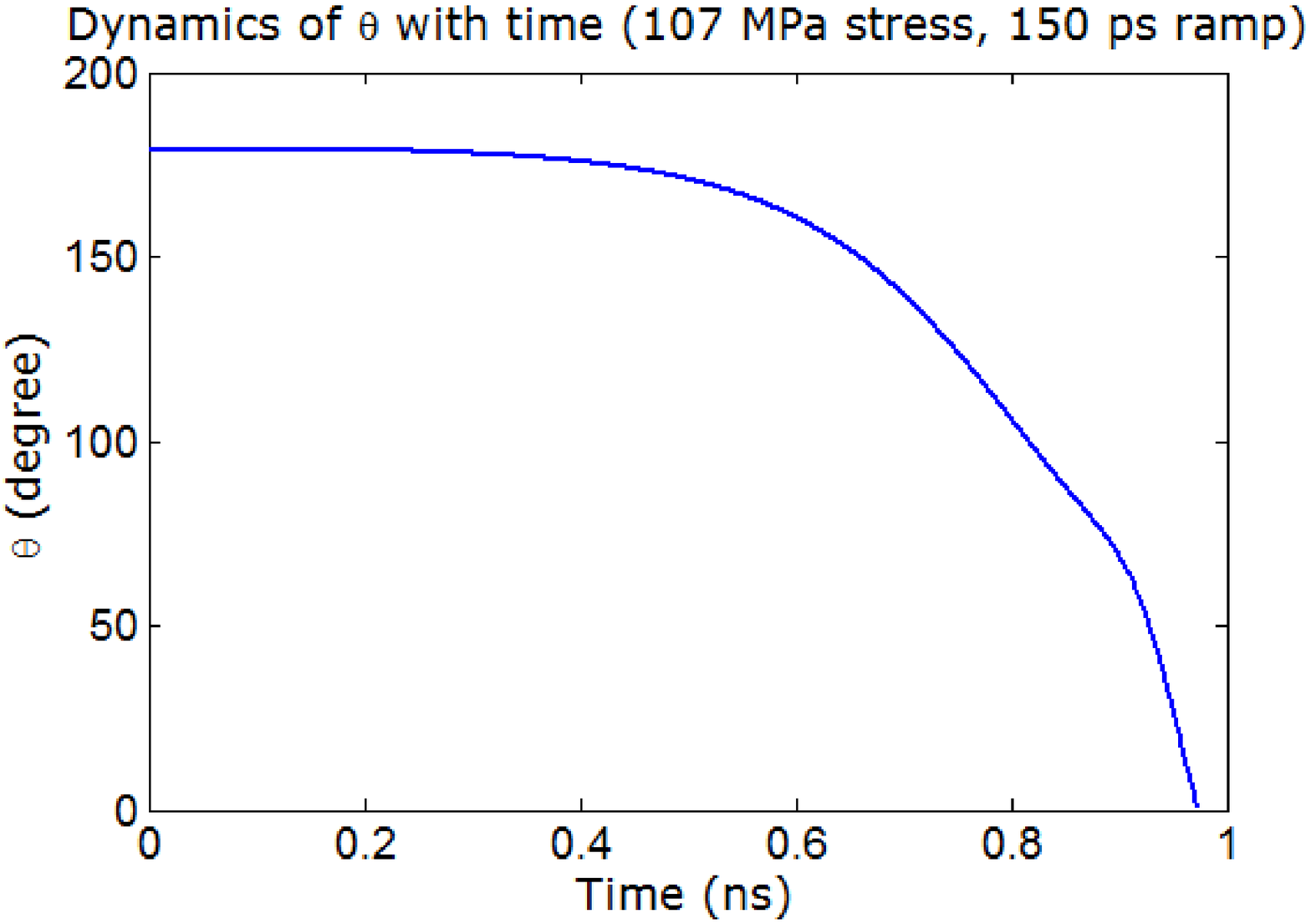}}
\subfigure[]{\label{fig:magnetization_rotation_nickel_107MPa_150ps}\includegraphics[width=3in]{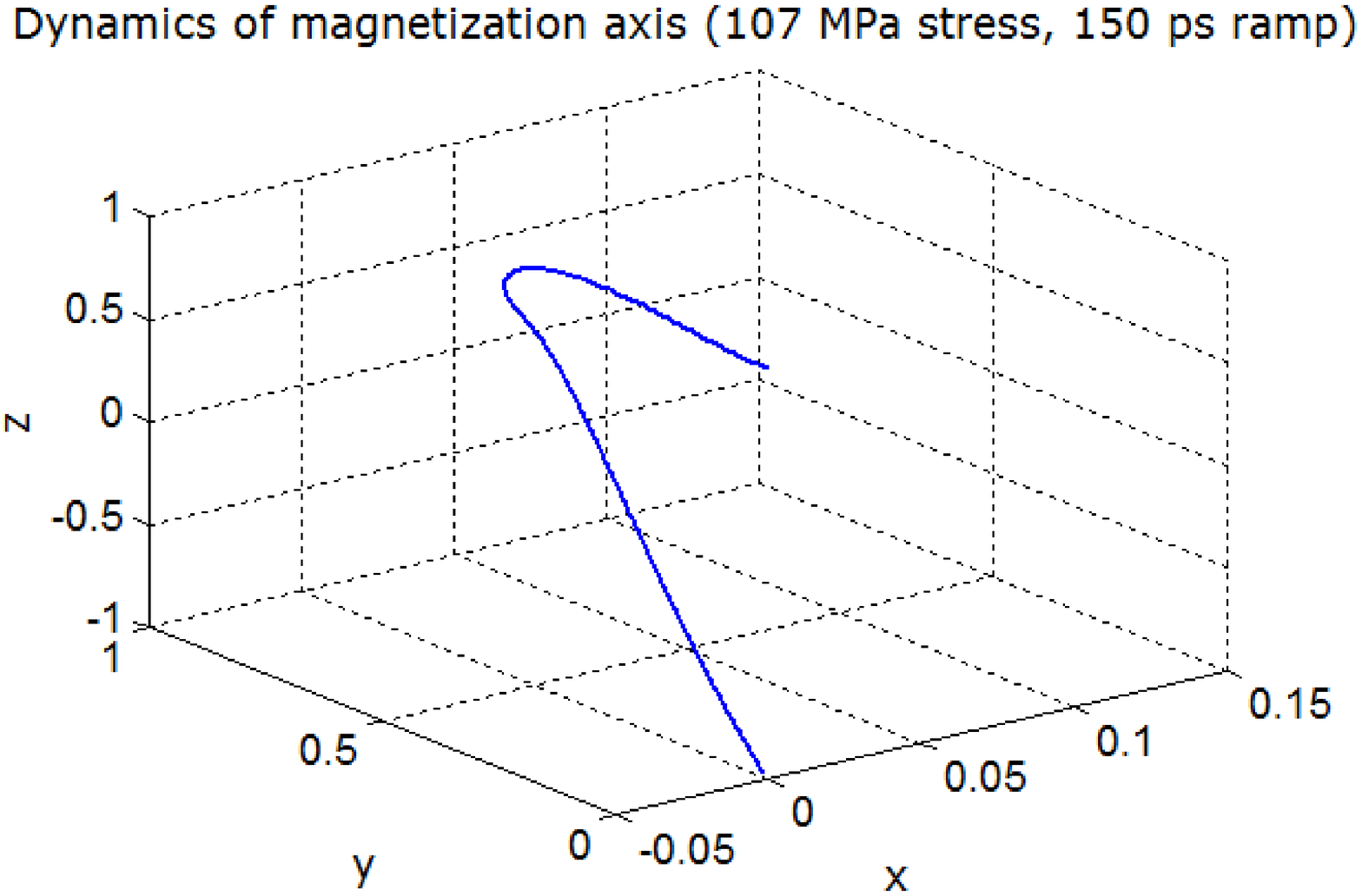}}
\caption{\label{fig:dynamics_nickel_107MPa_150ps} 
Magnetization dynamics in the nickel/PZT 
multiferroic nanomagnet. The stress is ramped 
up from 0 to 107 MPa in 150 ps: (a) polar angle $\theta$ versus time, and (b) the trajectory traced 
out by the tip of the magnetization vector while switching occurs.}
\end{figure}

\begin{figure}
\centering
\includegraphics[width=3in]{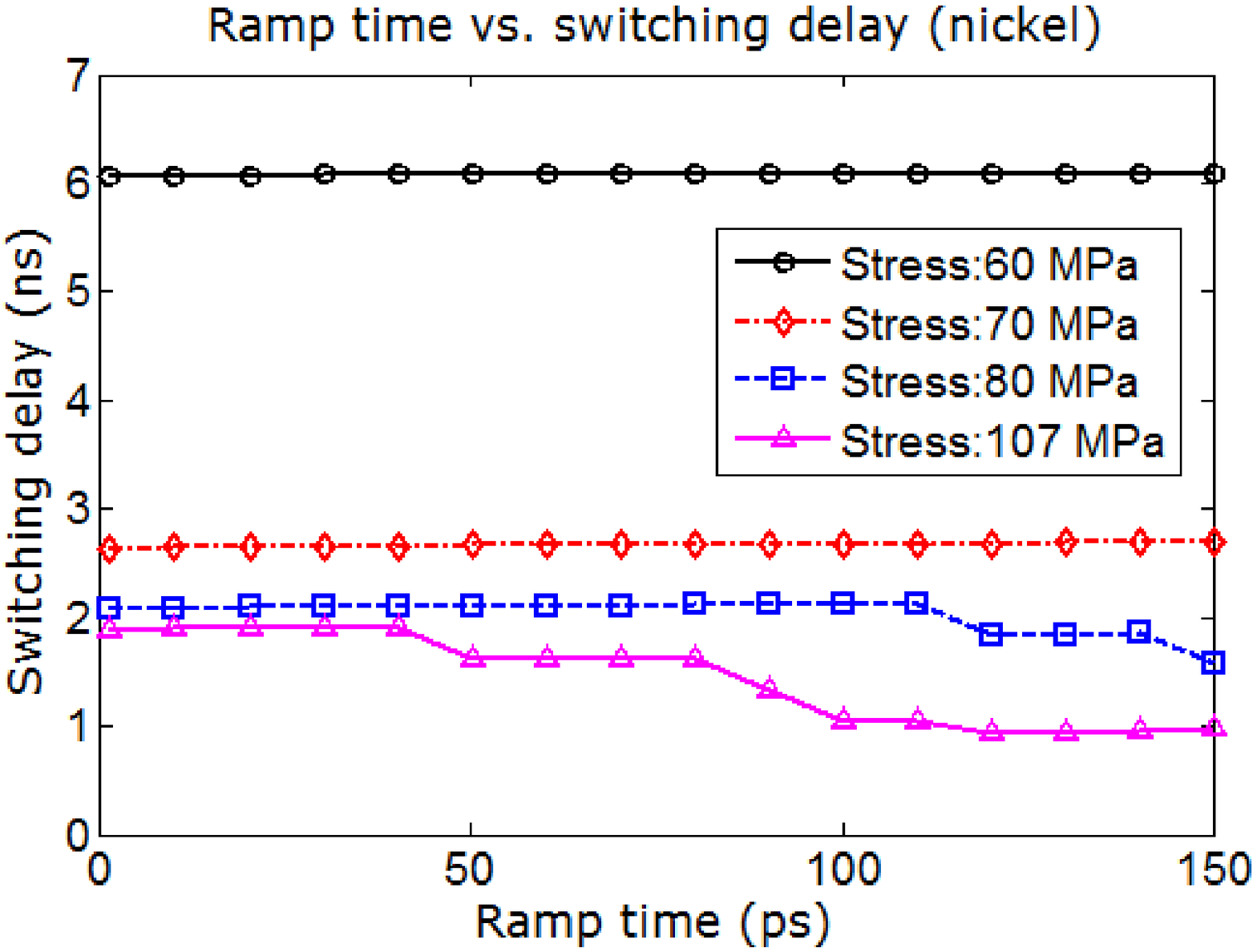}
\caption{\label{fig:delay_ramp_nickel} 
Switching delay of the nickel/PZT nanomagnet
as a function of the rise (or fall) time of the ramp, 
 with the magnitude of stress as a parameter.}
\end{figure}

\begin{figure}
\centering
\subfigure[]{\label{fig:theta_dynamics_nickel_60MPa_150ps}\includegraphics[width=3in]{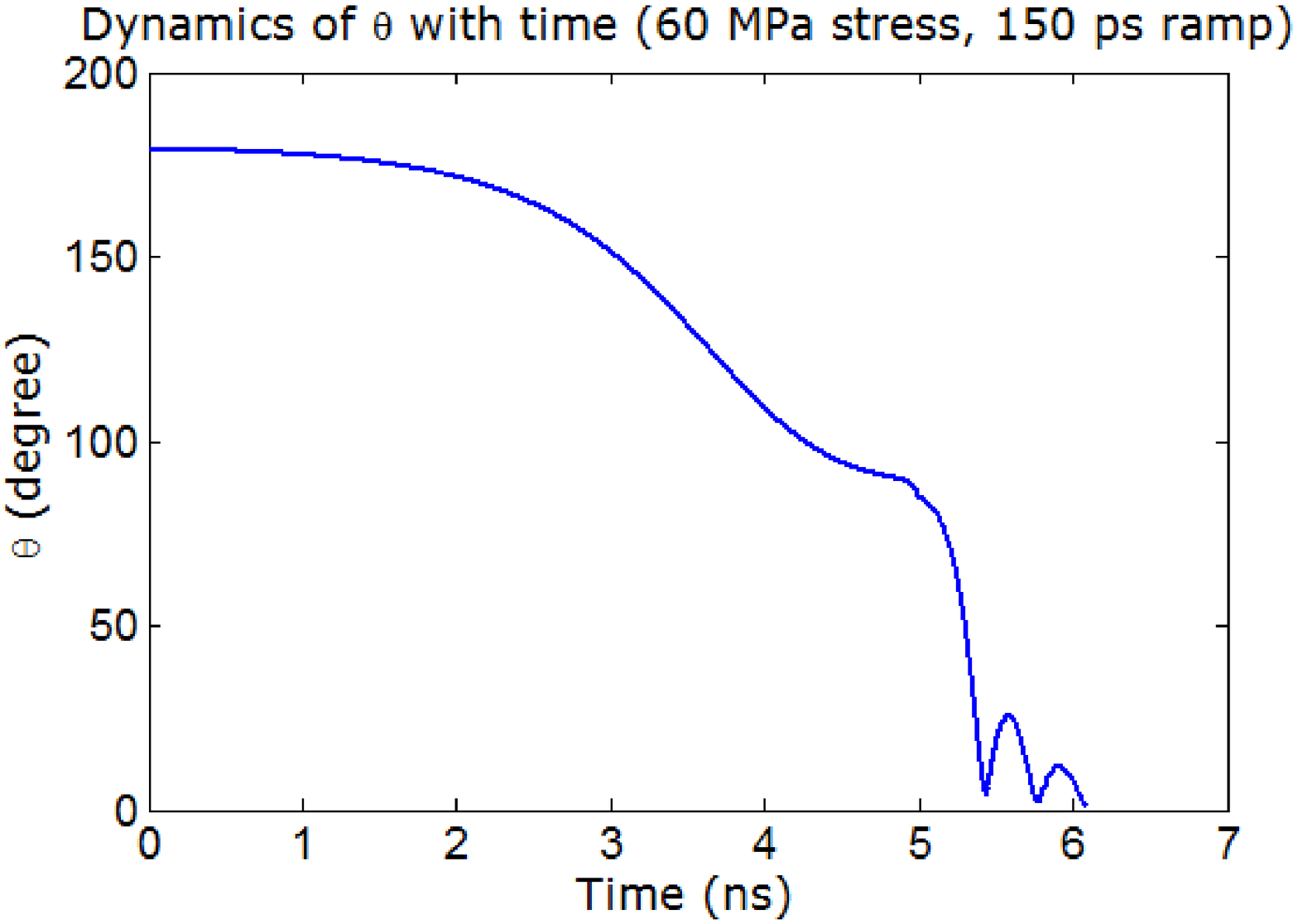}}
\subfigure[]{\label{fig:magnetization_rotation_nickel_60MPa_150ps}\includegraphics[width=3in]{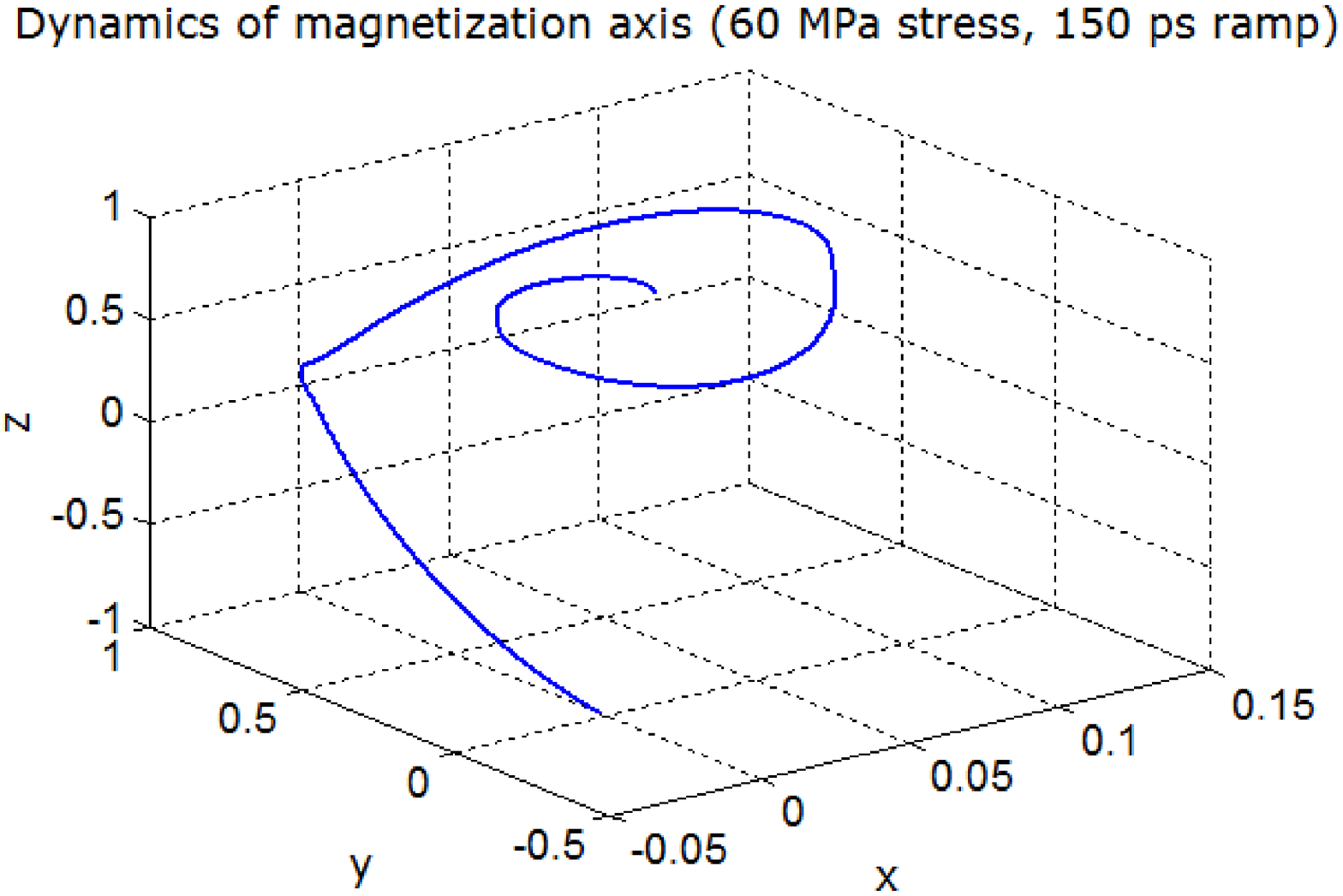}}
\caption{\label{fig:dynamics_nickel_60MPa_150ps} 
Magnetization dynamics in the nickel/PZT 
multiferroic nanomagnet when
the stress is ramped 
up linearly from 0 to 60 MPa in 150 ps - (a) The polar angle $\theta$ is plotted versus time. There is some 
ripple, but its amplitude 
is  reduced compared to the case when the rise (and fall) time is 1 ps. 
The switching delay in this case is about 6 ns.
(b) trajectory traced out by the tip of the magnetization vector while the switching occurs.}
\end{figure}
\begin{figure}
\centering
\subfigure[]{\label{fig:theta_dynamics_nickel_80MPa_150ps}\includegraphics[width=3in]{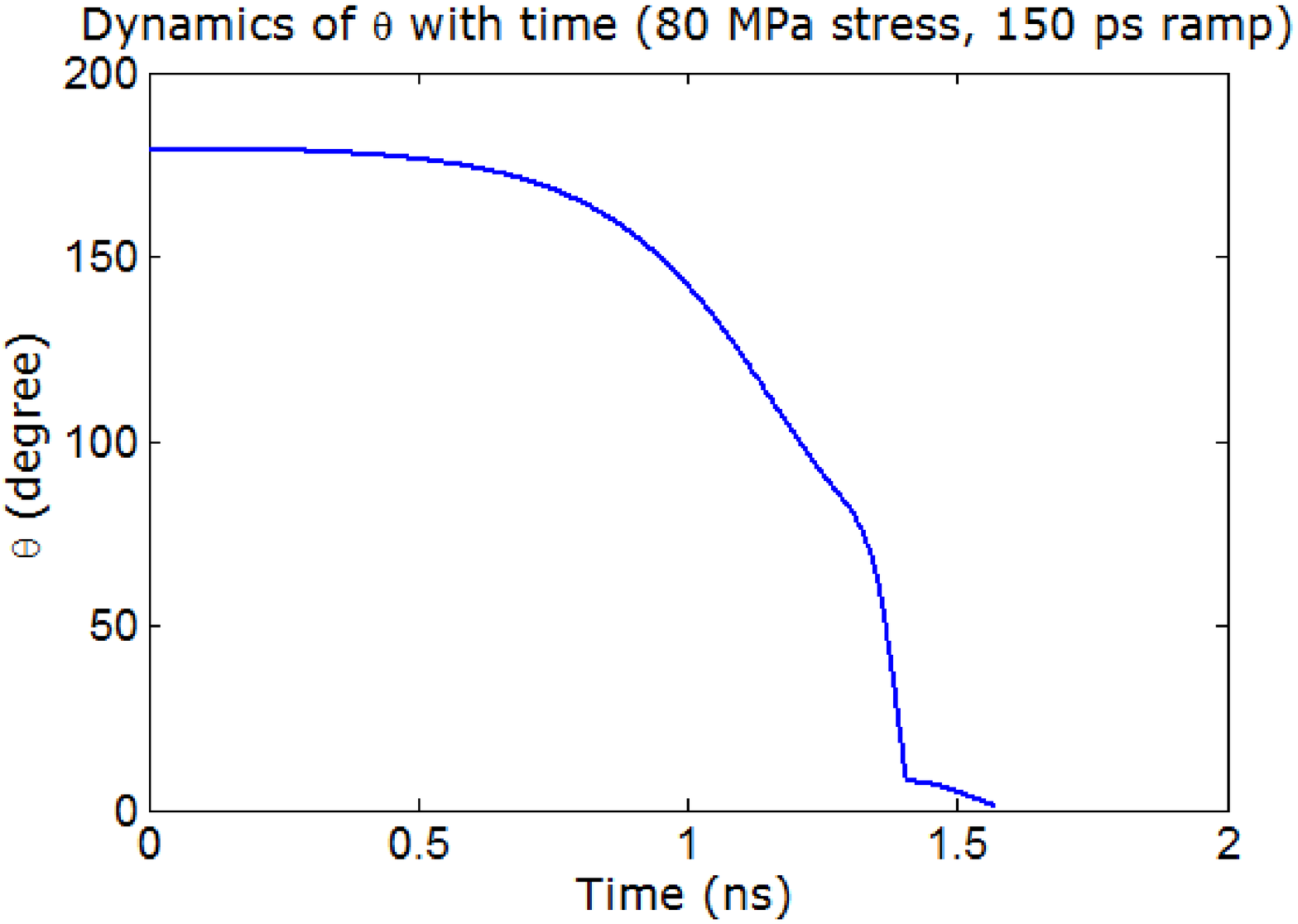}}
\subfigure[]{\label{fig:magnetization_rotation_nickel_80MPa_150ps}\includegraphics[width=3in]{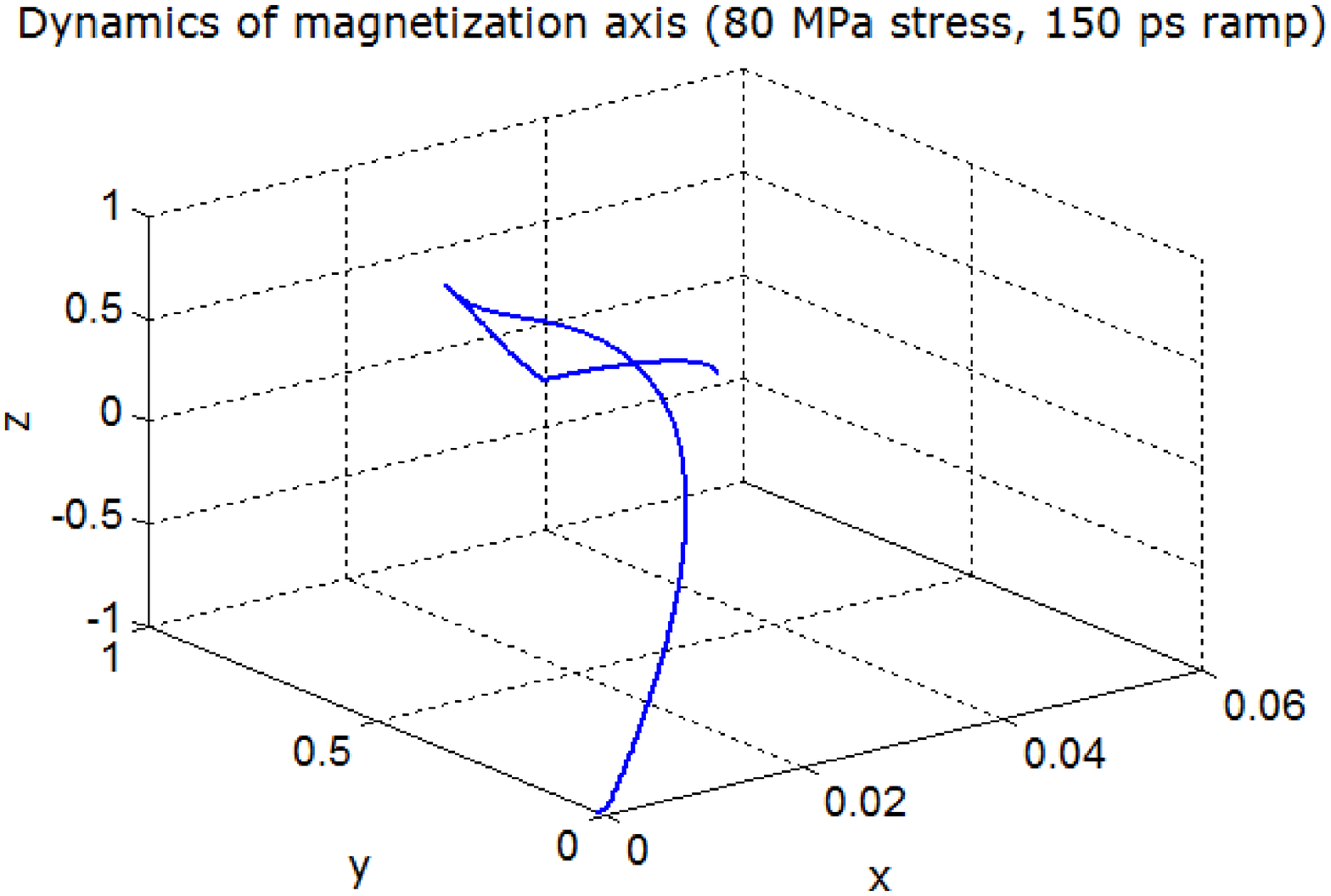}}
\caption{\label{fig:dynamics_nickel_80MPa_150ps} 
Magnetization dynamics in the nickel/PZT 
multiferroic nanomagnet when
the stress is ramped 
up linearly from 0 to 60 MPa in 150 ps - (a) The polar angle $\theta$ is plotted versus time. There is some 
ripple, but its amplitude 
is  reduced compared to the case when the rise (and fall) time is 1 ps. 
The switching delay in this case is about 1.6 ns.
(b) trajectory traced out by the tip of the magnetization vector while the switching occurs.}
\end{figure}

\begin{figure}
\centering
\includegraphics[width=3in]{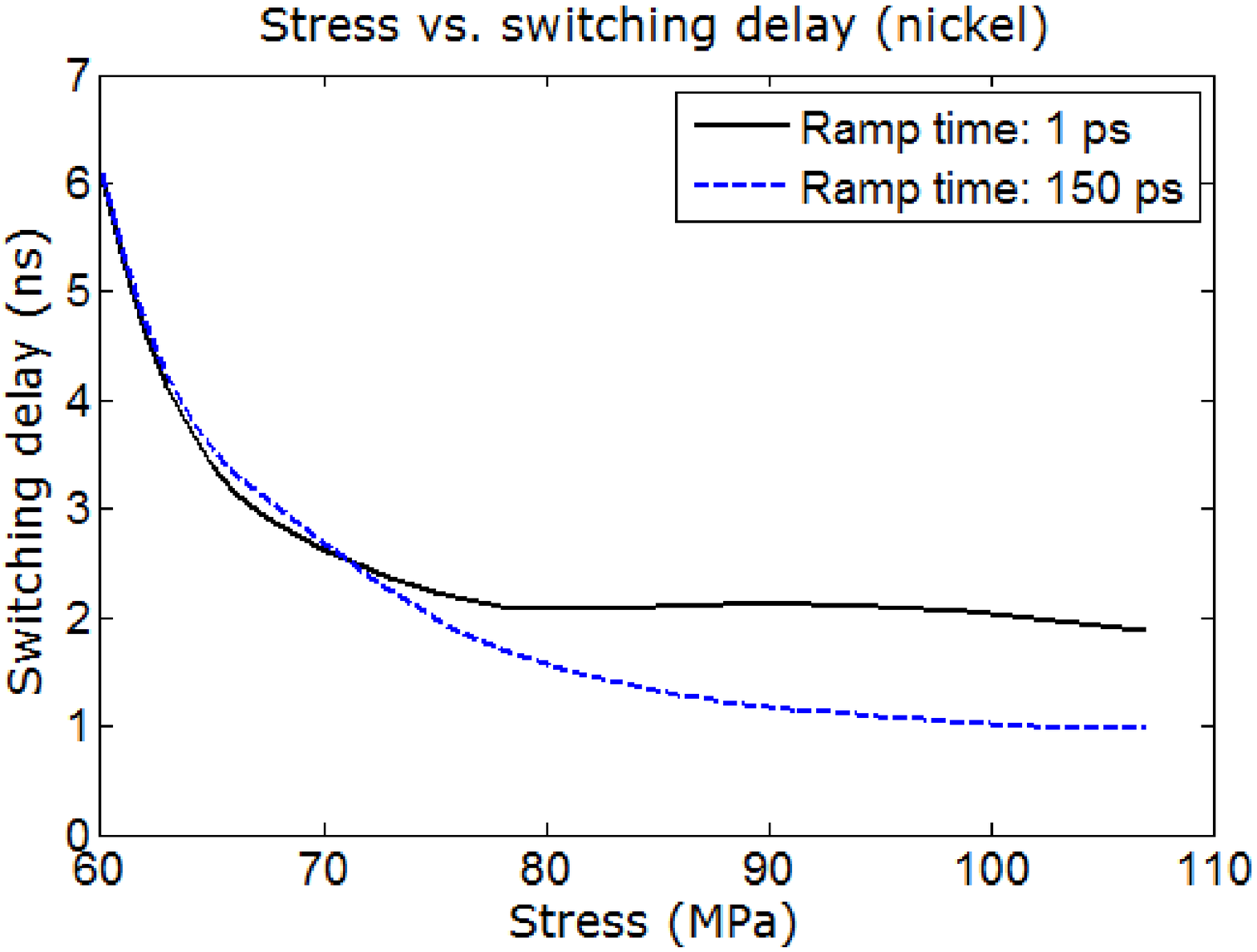}
\caption{\label{fig:delay_stress_nickel_ramp_time} 
Switching delay versus stress for the nickel/PZT
 multiferroic nanomagnet for two different ramp rise (and fall) times of 1 ps and 150 ps.}
\end{figure}

\begin{figure}
\centering
\includegraphics[width=3in]{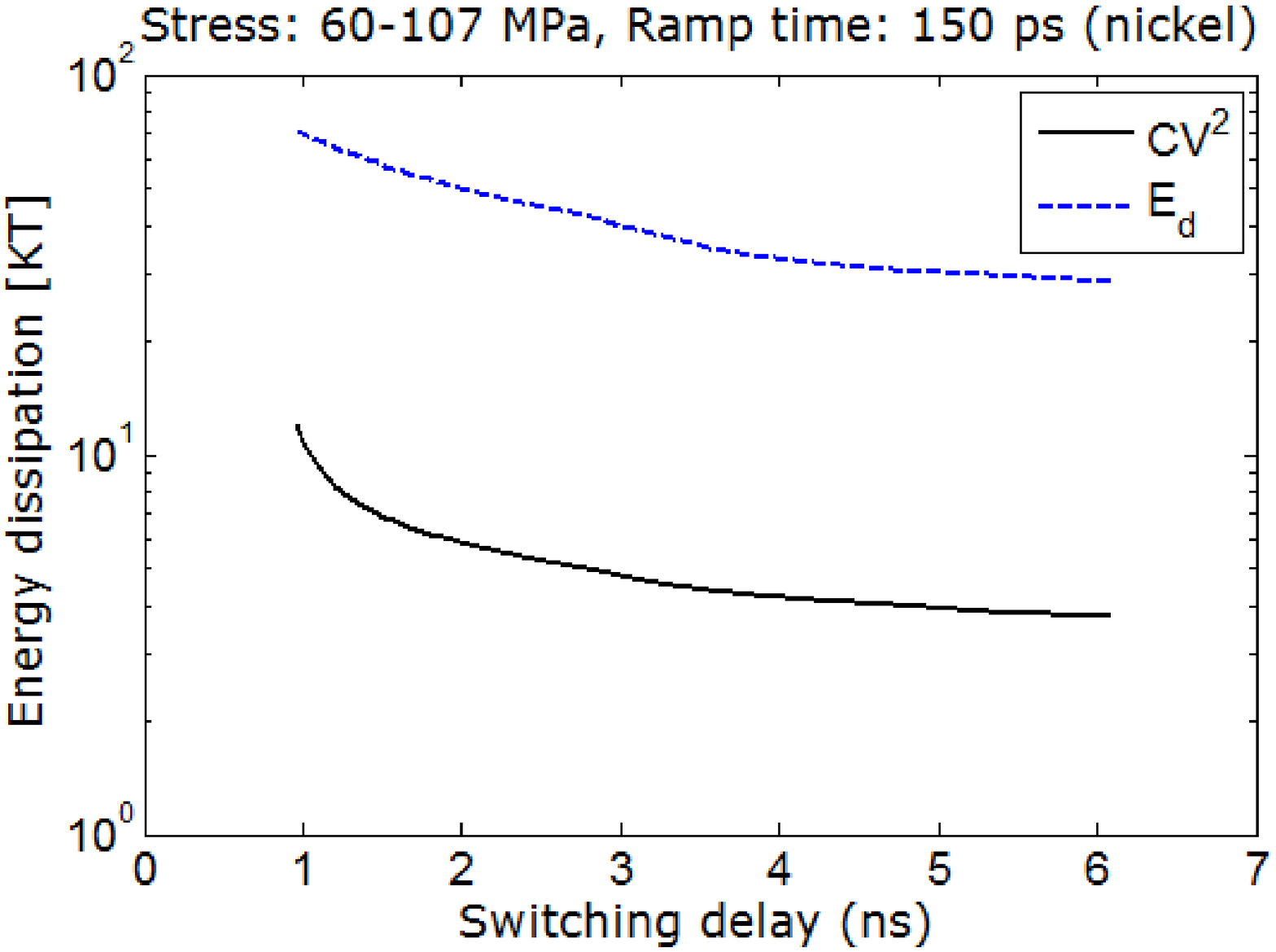}
\caption{\label{fig:delay_energy_nickel_stress} 
Energy dissipated in flipping the magnetization of the nickel/PZT multiferroic nanomagnet
as a function of switching delay
 for a ramp rise (and fall) time of 150 ps.
This range of switching delay corresponds to a stress range of 60 MPa to 107 MPa. The 
energy dissipated in the nanomagnet due to Gilbert damping and the energy dissipated in the external switching 
circuit (`$CV^2$') are shown separately.}
\end{figure}

\begin{figure}
\centering
\includegraphics[width=3in]{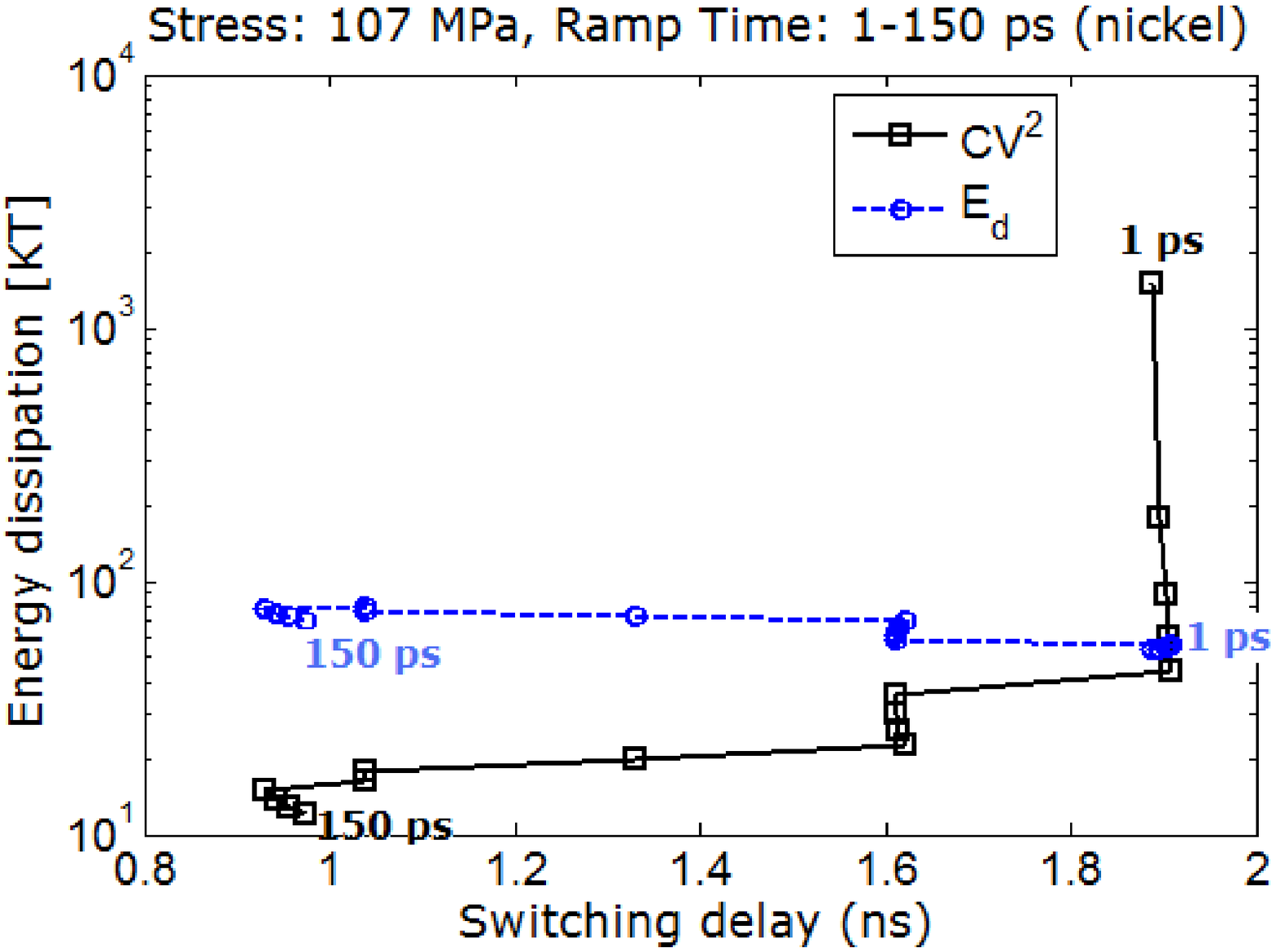}
\caption{\label{fig:delay_energy_nickel_ramp_time} For a fixed stress of 107 MPa,
energy dissipated in flipping the magnetization of the nickel/PZT nanomagnet 
as a function of switching delay when the latter is varied  
by continuously increasing the ramp's rise (and fall) time from 1-150 ps.}
\end{figure}

\subsection{Nickel}

Nickel has a negative magnetostrictive coefficient (see Table~\ref{tab:material_parameters}) 
so that a nickel/PZT multiferroic nanomagnet will require a \emph{tensile} stress to 
initiate rotation away from the easy axis.
However, the torque generated due to the change in stress with time 
(see Equations~\eqref{eq:theta_dynamics} and~\eqref{eq:phi_dynamics}) would be of same sign as that in
 the case of Terfenol-D since it depends on the \emph{product} of the magnetostrictive
 coefficient ($\lambda_s$) and the change in stress ($\partial \sigma$). For the dimensions of 
 the nanomagnet chosen, the minimum stress that we will need in a nickel/PZT multiferroic to switch
 is 57 MPa, while the maximum stress that can be generated by the 500 ppm strain in the PZT layer 
 is 107 MPa.

\subsubsection{Ramp rate and switching delay}

Just as in the case of Terfenol-D, Equations~\eqref{eq:theta_dynamics} and~\eqref{eq:phi_dynamics}
are solved numerically to find the values of $\theta(t)$ and $\phi(t)$ at any given instant $t$
for the nickel/PZT nanomagnet.
This yields the magnetization dyanmics under various stresses and ramp rates.

\paragraph{Fast ramp:}

Figure~\ref{fig:dynamics_nickel_107MPa_1ps} shows the magnetization dynamics of a nickel/PZT multiferroic 
nanomagnet when the stress is ramped up linearly in time from 0 to the maximum possible value 
of 107 MPa in 1 ps.  The in-plane and out-of-plane dynamics of the magnetization vector are very similar to 
the case of Terfenol-D, except now we see a more ripples since there is more out-of-plane precession 
as can be seen in Figure~\ref{fig:dynamics_nickel_107MPa_1ps}(b). Nickel shows more ripples and more
precession because it has a smaller Gilbert damping constant than Terfenol-D. Consequently, the precessional motion is less damped.

\paragraph{Slow ramp:}

Figure~\ref{fig:dynamics_nickel_107MPa_150ps}(b)
shows that a slow ramp rate (150 ps rise (and fall) time) quenches
the precession of the magnetization vector. 
This happens because a slower ramp causes less out-of-plane excursion of the magnetization
vector and hence less precessional motion. Note that contrary to expectation, the slow ramp (150 ps rise (and fall) time)
 completes the switching 
in 0.95 ns, while the fast ramp (1 ps rise (and fall) time) takes twice the amount of time, i.e. 1.9 ns! This happens
because of the ripples and ringing which have a deleterious effect on switching speed.
Note that both ramp rates make $\theta$ come very close to 1$^{\circ}$ in about 0.9 ns, but the faster 
ramp rate causes the magnetization vector to pull back from the final destination and vacillate before 
reaching the final destination. This causes the ringing which prolongs the switching duration and increases the delay. 

Figure~\ref{fig:delay_ramp_nickel} shows switching delay as a function of 
the ramp's rise (and fall) time for various stresses. At higher stress levels (80 - 107 MPa), the switching delay actually
{\it decreases} with increasing rise (and fall) time (a counter-intuitive result) because of the 
ripples that are generated when the stress is ramped up to high values very fast. Thus, the out-of-plane
dynamics of the magnetization vector plays a vital role in switching.

For lower stress levels (60 -70 MPa), the switching delay increases slightly with increasing rise (or fall) time. 
Low stresses do not cause significant out-of-plane excursion of the magnetization vector leading to
precessional motion. Therefore, even a fast ramp does not cause too many ripples and delay the switching when 
the stress
is sufficiently low. In that case, a slower ramp will lead to slightly slower switching and that is 
what we observe. 

Strangely, at slow ramp rates (e.g. rise (and fall) time = 150 ps) and low stress levels, 
a slightly lower stress can cause considerably more precession than
a slightly higher stress which leads to very strong dependence of switching delay on the magnitude of 
stress at low stress levels. In order to illustrate this, we plot the switching dynamics for 150 ps rise (and fall) time in
Figures~\ref{fig:dynamics_nickel_60MPa_150ps} and~\ref{fig:dynamics_nickel_80MPa_150ps}, 
for stresses of 60 MPa and 80 MPa. There are ripples and
considerable precessional motion when the stress is 60 MPa and virtually no ripple and very little 
precessional motion when the stress is 80 MPa. Thus, by increasing the stress slightly from 60 MPa
to 80 MPa, the switching time can be reduced by a factor of four -- from slightly above 6 ns to slightly
above 1.5 ns. 

The strong dependence of switching delay on stress at low stress levels is illustrated in 
Figure~\ref{fig:delay_stress_nickel_ramp_time} which plots switching delay as a function of 
stress for two different rise (and fall) times of 1 ps and 150 ps. Notice that switching delay increases
rapidly with decreasing stress in the interval [60 MPa, 70 MPa] but much less rapidly at higher 
stress levels exceeding 80 MPa. This is purely a consequence of the complex out-of-plane 
dynamics of the magnetization vector. This shows that any analysis which ignores the out-of-plane 
dynamics, and tacitly assumes that the motion of the magnetization vector will be always constrained to 
the plane of the nanomagnet since $N_{d-xx} \gg N_{d-yy}, N_{d-zz}$, will not only be quantitatively wrong,
but qualitatively wrong as well.

The cross over between the two curves in Figure~\ref{fig:delay_stress_nickel_ramp_time} was explained 
in the context of Terfenol-D and is not repeated here. This too is a consequence of the out-of-plane 
dynamics.

\subsubsection{Switching delay and energy dissipation}

Figure~\ref{fig:delay_energy_nickel_stress} shows the energy dissipated in flipping the 
magnetization of the nickel/PZT multiferroic nanomagnet as a function of the 
switching delay. The latter is varied by varying the applied stress between 60
and 107 MPa with fixed rise (and fall) time of 150 ps for the stress ramp.
The energy dissipated internally in the nanomagnet due to Gilbert damping and the `$CV^2$' energy
dissipated in the external circuit are shown separately.
Both dissipation components decrease smoothly with increasing switching delay, implying that 
the average power dissipated during switching decreases rapidly with increasing delay. 
Both tend to saturate as the switching delay becomes longer.

In Figure~\ref{fig:delay_energy_nickel_stress}, note that the `$CV^2$' energy dissipated in the 
switching circuit is 1-2 orders of magnitude higher for nickel than for Terfenol-D for
the same switching delay. Since the 
voltage $V$ is proportional to stress, the `$CV^2$' energy is quadratically proportional to 
stress. Since the magnetostrictive coefficient of Terfenol-D is considerably higher than that of
nickel, Terfenol-D requires much less stress to generate the same stress anisotropy energy and 
hence requires much less stress to switch. This results in a significant reduction of 
`$CV^2$' energy dissipation in the case of Terfenol-D.

The total energy dissipation is however dominated by the energy $E_d$ dissipated internally in the nanomagnet 
due to Gilbert damping. This energy is actually smaller in nickel than in Terfenol-D, because the 
Gilbert damping constant of nickel is more than twice smaller than that of Terfenol-D. As long
as we are switching adiabatically, $E_d$ will be the primary source of dissipation, and in that case,
the material with the lower Gilbert damping constant will be superior since it will reduce total 
dissipation. {\it The situation may change completely if we are switching abruptly}. In that 
case, the `$CV^2$' energy may very well be the major component of dissipation. If that happens, then the 
material with the larger magnetostrictive coefficient will be better since it will need less stress 
to switch and hence less voltage and less `$CV^2$' dissipation. In other words, nickel is better than Terfenol-D 
(from the perspective of energy dissipation) when switching 
is adiabatic, but the opposite may very well be true when the switching is abrupt.

If speed is the primary concern, then what matters is the product of the magnetostrictive 
coefficient and the Young's modulus. Since the maximum strain that can be generated in the 
magnetostrictive layer is fixed and 
determined by the PZT layer, the maximum stress anisotropy energy that can be generated in the magnet
depends on the aforesaid
product. The higher the stress anisotropy energy is, the faster will be the switching.
Since the product is higher for Terfenol-D than for nickel, the Terfenol-D/PZT multiferroic switches 
faster. It can be switched in sub-ns by the maximum strain generated in the PZT layer, but that same 
strain cannot switch a nickel/PZT multiferroic in less than 1 ns.

Figure~\ref{fig:delay_energy_nickel_ramp_time} shows the energy dissipated 
as a function of switching delay when the latter is varied by varying the 
ramp's rise (and fall) time between 1 ps and 150 ps, while holding the stress constant at 107 MPa.
The `$CV^2$' component decreases with increasing rise (and fall) time because switching becomes 
increasingly adiabatic. However, the dependence of $E_d$ on switching delay is more complicated. When the rise 
and fall of the ramp is fast, ripples appear. That causes the energy dissipation to increase with increasing 
delay contrary to expectation. This corresponds to the segment of the curve between switching delays of 0.9 to 1.9 ns
which corresponds to rise times between 1 and 120 ps. At slower rise times between 120 and 150 ps, the ripples
subside and the energy dissipation $E_d$ decreases with increasing delay. This corresponds to the segment between 0.9 ns
and 1 ns switching delay.

\begin{figure}
\centering
\subfigure[]{\label{fig:theta_dynamics_cobalt_104d5MPa_1ps}\includegraphics[width=3in]{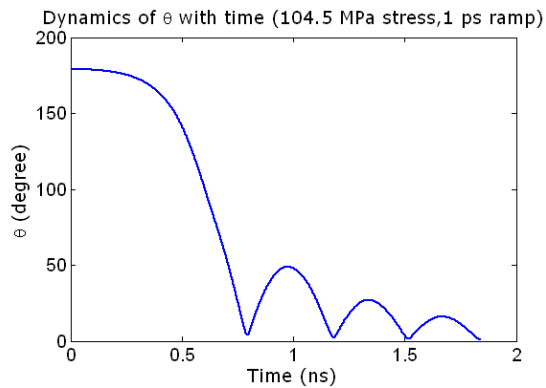}}
\subfigure[]{\label{fig:magnetization_rotation_cobalt_104d5MPa_1ps}\includegraphics[width=3in]{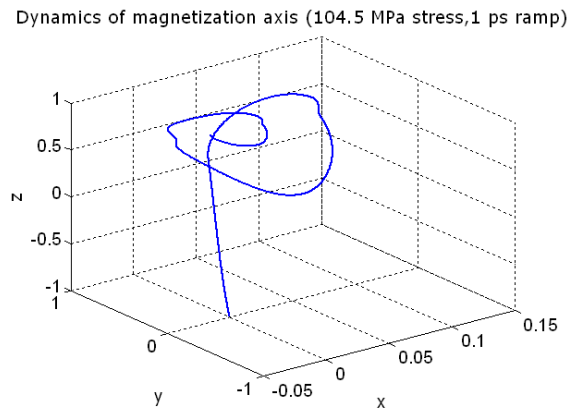}}
\caption{\label{fig:dynamics_cobalt_104d5MPa_1ps} 
Magnetization dynamics in the cobalt/PZT multiferroic 
nanomagnet. The stress is ramped 
up from 0 to 104.5 MPa in 1 ps: (a) polar angle $\theta$ versus time, and (b) the trajectory traced out by the 
tip of the magnetization vector in three-dimensional space while switching occurs, i.e. during 
the time $\theta$ changes from 179$^{\circ}$ to 1$^{\circ}$.}
\end{figure}
\begin{figure}
\centering
\subfigure[]{\label{fig:theta_dynamics_cobalt_104d5MPa_150ps}\includegraphics[width=3in]{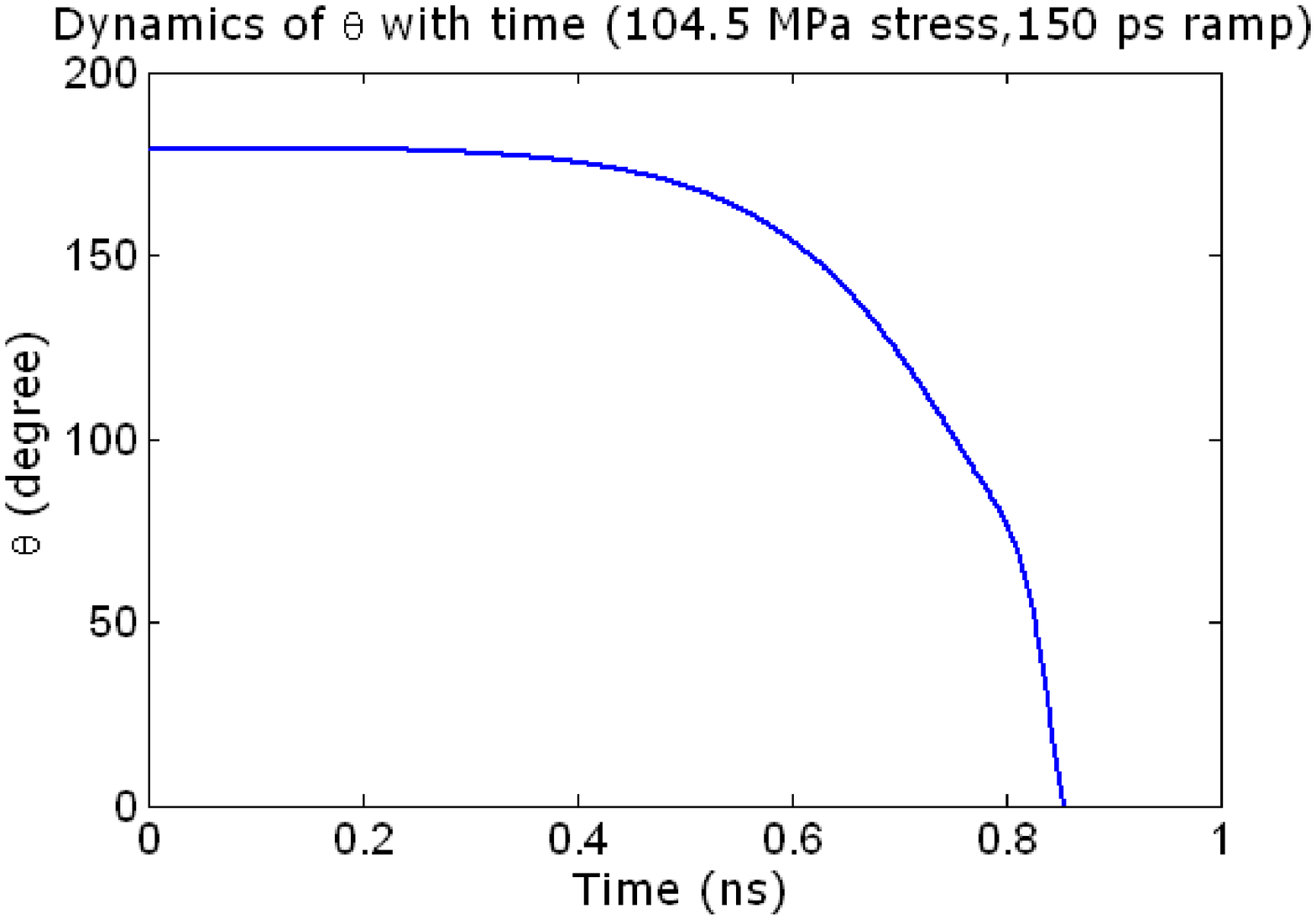}}
\subfigure[]{\label{fig:magnetization_rotation_cobalt_104d5MPa_150ps}\includegraphics[width=3in]{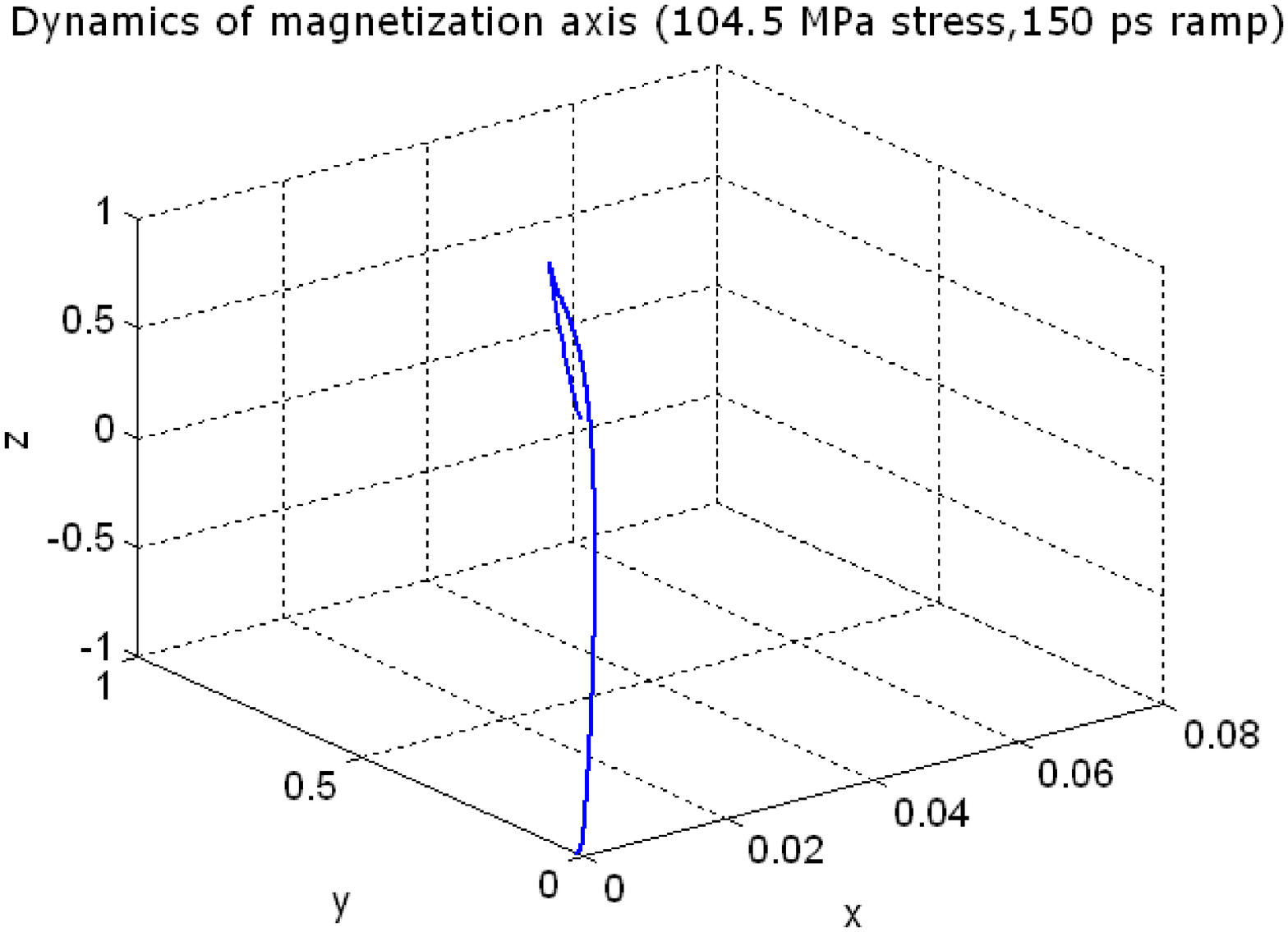}}
\caption{\label{fig:dynamics_cobalt_104d5MPa_150ps} 
Magnetization dynamics in the cobalt/PZT 
multiferroic nanomagnet. The stress is ramped 
up from 0 to 104.5 MPa in 150 ps: (a) polar angle $\theta$ versus time, and (b) the trajectory traced 
out by the tip of the magnetization vector while switching occurs.}
\end{figure}

\begin{figure}
\centering
\includegraphics[width=3in]{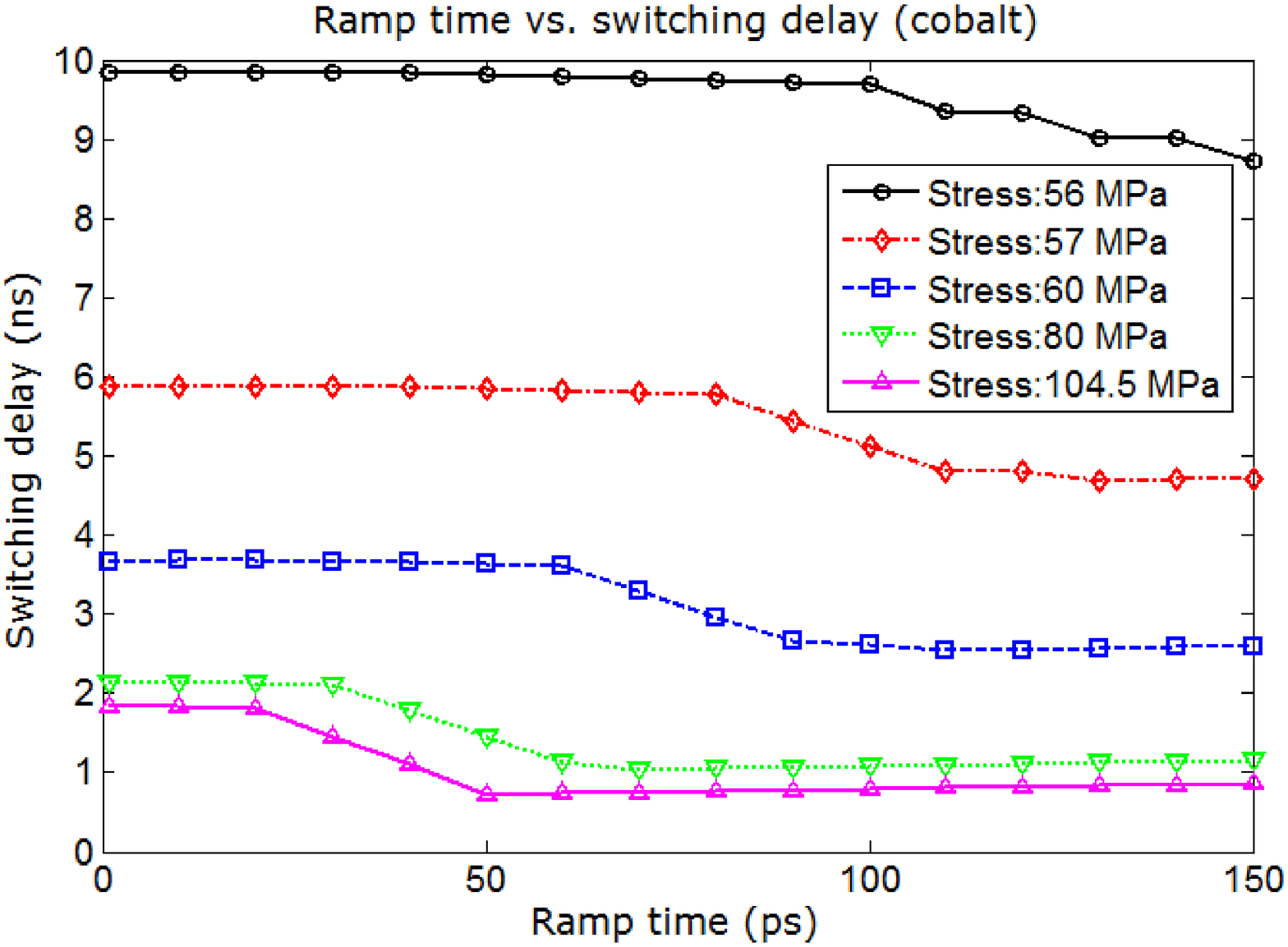}
\caption{\label{fig:delay_ramp_cobalt} Switching delay of the cobalt/PZT nanomagnet
as a function of the rise (or fall) time of the ramp, 
 with the magnitude of stress as a parameter.}
\end{figure}

\begin{figure}
\centering
\includegraphics[width=3in]{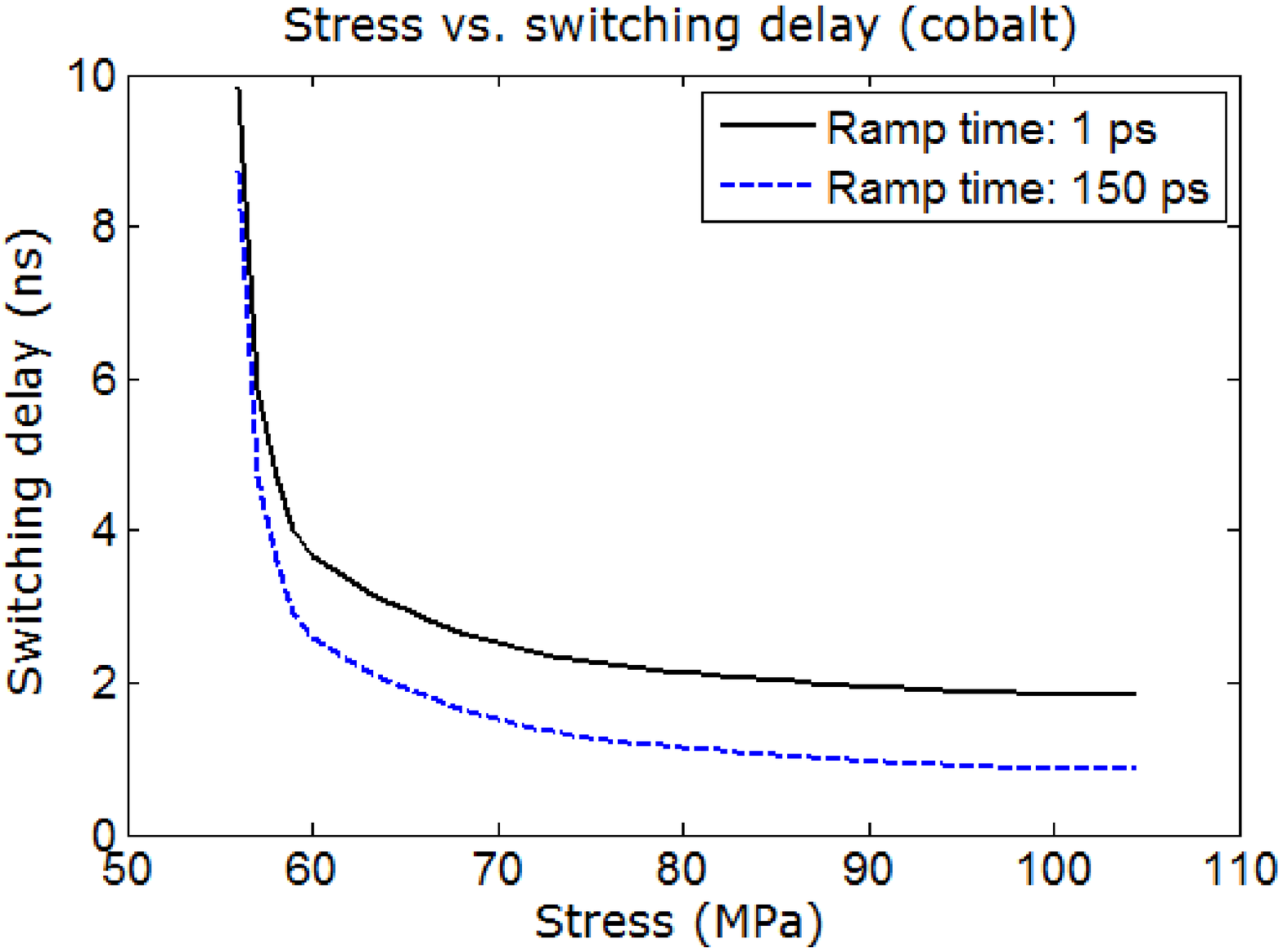}
\caption{\label{fig:delay_stress_cobalt_ramp_time} 
Switching delay versus stress for the cobalt/PZT
 multiferroic nanomagnet for two different ramp rise (and fall) times of 1 ps and 150 ps.}
\end{figure}

\begin{figure}
\centering
\includegraphics[width=3in]{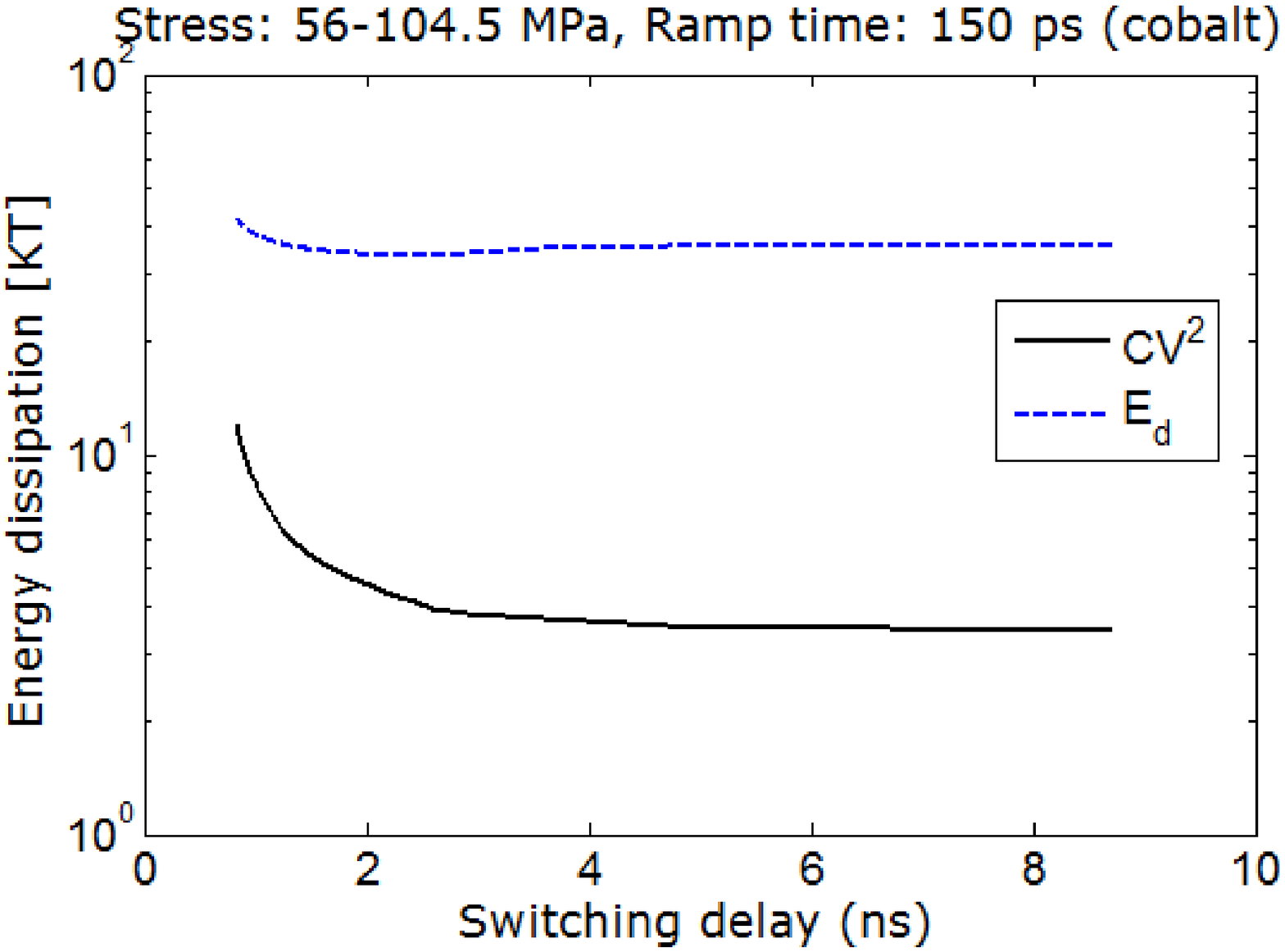}
\caption{\label{fig:delay_energy_cobalt_stress} 
Energy dissipated in flipping the magnetization of the cobalt/PZT multiferroic nanomagnet
as a function of switching delay
 for a ramp rise (and fall) time of 150 ps.
This range of switching delay corresponds to a stress range of 56 MPa to 104.5 MPa. The 
energy dissipated in the nanomagnet due to Gilbert damping and the energy dissipated in the external switching 
circuit (`$CV^2$') are shown separately.}
\end{figure}

\begin{figure}
\centering
\includegraphics[width=3in]{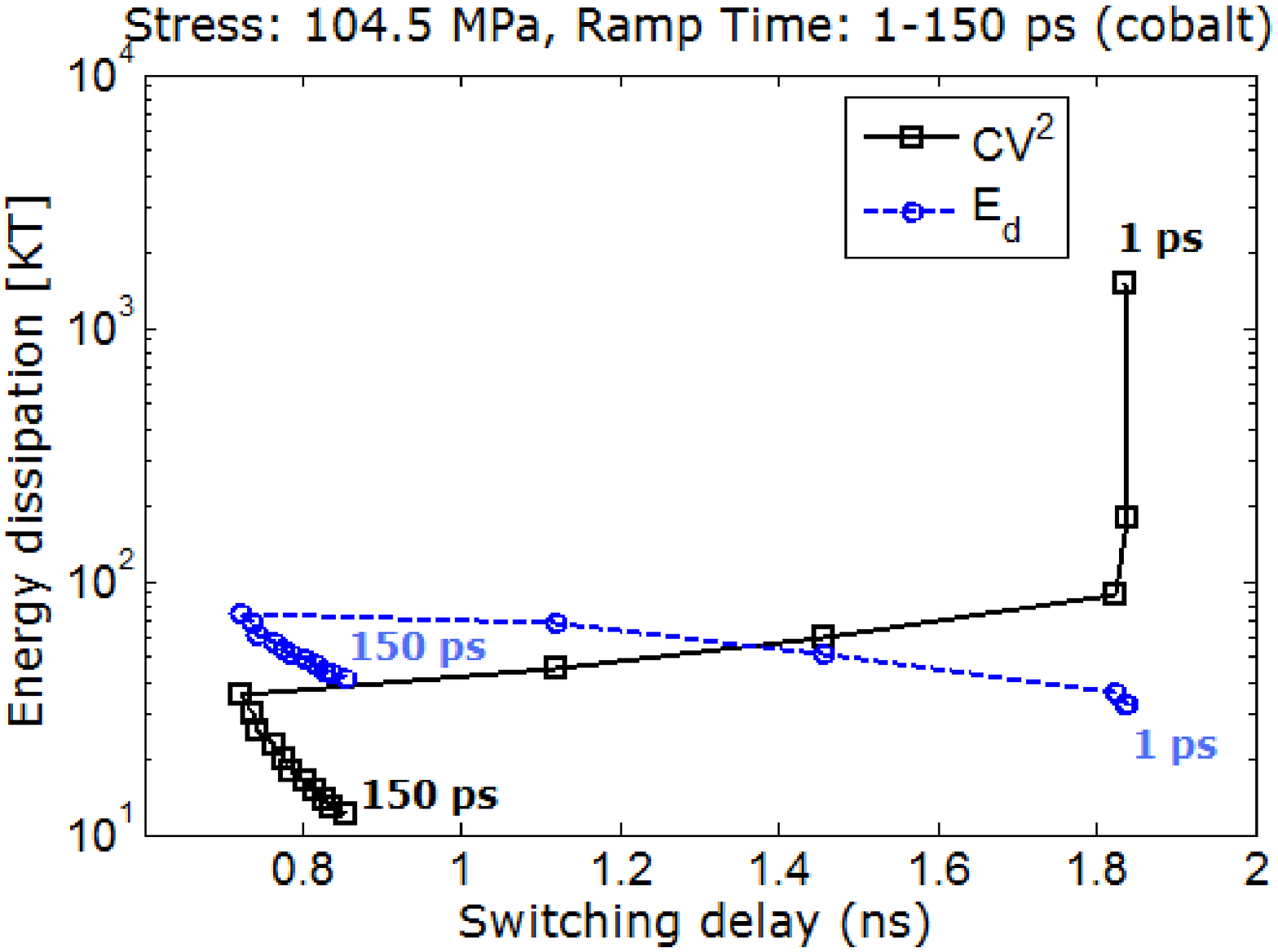}
\caption{\label{fig:delay_energy_cobalt_ramp_time} 
For a fixed stress of 104.5 MPa,
energy dissipated in flipping the magnetization of the cobalt/PZT nanomagnet 
as a function of switching delay when the latter is varied  
by continuously increasing the ramp's rise time from 1-150 ps.}
\end{figure}

\subsection{Cobalt}

Cobalt has a negative magnetostrictive coefficient that is similar to nickel's. 
Therefore, we will 
 need a \emph{tensile} stress to initiate magnetization rotation away from the easy axis.
Its Gilbert damping constant is however smallest among the three
materials considered (see Table~\ref{tab:material_parameters})
and hence we expect it to be least dissipative. For the dimensions of 
 the nanomagnet chosen, the minimum stress that we will need in a cobalt/PZT multiferroic to switch
 is 56 MPa, while the maximum stress that can be generated by the 500 ppm strain in the PZT layer 
 is 104.5 MPa.
 
\subsubsection{Ramp rate and switching delay}

As before, Equations~\eqref{eq:theta_dynamics} and~\eqref{eq:phi_dynamics}
are solved numerically to find the values of $\theta(t)$ and $\phi(t)$ at any given instant $t$
for the cobalt/PZT nanomagnet.

\paragraph{Fast ramp:}

Figure~\ref{fig:dynamics_cobalt_104d5MPa_1ps} shows the magnetization dynamics of a cobalt/PZT multiferroic 
nanomagnet when the stress is ramped up linearly in time from 0 to the maximum possible value 
of 104.5 MPa in 1 ps.  The in-plane and out-of-plane dynamics of the magnetization vector are very similar to 
the case of Terfenol-D or nickel, except now we see even more ripples and even more out-of-plane 
precession. Cobalt shows the most prominent ripples because it has the smallest Gilbert damping constant among all three ferromagnets
considered. Hence, the precessional motion is least damped.

\paragraph{Slow ramp:}
Figure~\ref{fig:dynamics_cobalt_104d5MPa_150ps} 
shows that a slow ramp rate (150 ps rise (and fall) time) quenches
the precession of the magnetization vector. 
This happens because a slower ramp causes less out-of-plane excursion of the magnetization
vector and hence less precessional motion. This reduces the switching delay from 
1.6 ns in the case of 1 ps rise (and fall) time to 0.85 ns for 150 ps rise (and fall) time.

Figure~\ref{fig:delay_ramp_cobalt}  shows switching delay as a function of 
the ramp's rise (and fall) time for various stresses. At all stress levels -- not just at high stress levels 
as in the case of nickel -- the switching delay 
decreases with increasing rise (and fall) time  because of the 
ripples and the underlying precession that are generated when the stress is ramped up very fast. 
Unlike in the case of nickel, this happens even at small stress levels in cobalt since cobalt has the
smallest Gilbert damping constant among the three and is hence most susceptible to precession.

\subsubsection{Switching delay and energy dissipation}

Figure~\ref{fig:delay_stress_cobalt_ramp_time} shows the dependence of switching delay 
on stress for two different ramp rise (and fall) times of 1 ps and 150 ps. 
As expected, the switching delay has a very weak dependence on the rise (and fall) time at small
stress levels when not too much ringing  occurs. However, at high stress levels, the 
switching delay increases considerably for the shorter ramp time since that 
causes prolonged ringing associated with precessional motion. This is a more pronounced 
effect for cobalt than for nickel or Terfenol-D since cobalt has the smallest Gilbert damping 
constant among all three that makes it particularly susceptible to precessions.

Figure~\ref{fig:delay_energy_cobalt_stress} shows the energy dissipated in flipping the magnetization
as a function of switching delay when the latter is varied by varying the stress between 56 MPa and 
the maximum possible 104.5 MPa. Both the internal energy dissipation $E_d$ and the energy dissipated in
the switching circuit `$CV^2$' decreases with increasing delay showing that the average power dissipated 
decreases quite rapidly with increasing delay. Overall, the energy dissipation is slightly less here than 
in nickel because cobalt has the lower Gilbert damping constant
and slightly higher saturation magnetization (see Equation (\ref{eq:Ed_dissipation}). 
Once again, the dissipated energies saturate at long delays.

Figure~\ref{fig:delay_energy_cobalt_ramp_time} plots the energy dissipations as a
function of the switching delay when the latter is varied by monotonically varying the 
rise time of the ramp from 1 to 150 ps, while holding the stress constant at 104.5 MPa.
 For the $E_d$ plot, there are  two segments. In the first segment spanning the range of
 switching delay between 0.75 ns and 1.8 ns, the ramp's rise (and fall) time was varied from 
 1 ps to 50 ps. In this interval, the switching delay decreases with increasing rise (and fall) time 
 because of the ripple effect
 (see Figure~\ref{fig:delay_ramp_cobalt} and the 104.5 MPa stress curve). In this interval, 
 the energy dissipated internally in the magnet
 $E_d$ goes up with increasing ramp time since we are switching faster, but the `$CV^2$' energy
 dissipated in the external circuit goes down since switching is becoming more adiabatic (rise time is longer)$\footnote{Note
that adibaticity of switching is not determined by the switching delay, but by the rise (and fall) time of the ramp.}$.
 The second segment spanning the switching delay range between 0.75 and 0.85 ns corresponds to
 rise (and fall) time between 50 ps and 150 ps. In this range of rise (and fall) times, the switching delay increases
 with increasing rise (and fall) time because the ripples subside 
 (see Figure~\ref{fig:delay_ramp_cobalt} again and the 104.5 MPa stress curve). In this range,
 $E_d$ goes down with increasing ramp time since we are switching slower.
 The `$CV^2$' energy dissipation of course always decreases with increasing rise (and fall) time since the 
 switching becomes increasingly adiabatic.

\section{\label{sec:discussions}Discussions}
We have analyzed the switching dynamics in a multiferroic nanomagnet consisting of a PZT layer and 
a magnetostrictive layer subjected to time-varying stress. The stress is ramped up linearly in time 
with different rates or rise (and fall) times. Three different materials (Terfenol-D, nickel, cobalt) were considered
for the magnetostrictive layer. They show different behavior because of different 
material parameters (Youngs' modulus, magnetostrictive coefficient and Gilbert damping).

 For the type of magnets chosen (materials and dimensions), the minimum switching delay that we have found 
 is 252 ps obtained with Terfenol-D 
 using a rise (and fall) time of 60 ps for a compressive stress of
 40 MPa. The corresponding `$CV^2$' energy dissipation in the switching 
 circuit is 30 kT and the energy dissipated internally in the nanomagnet 
 due to Gilbert damping is $E_d$ = 777 kT. For nickel,
 the minimum switching delay is 930 ps obtained with a stress of 107 MPa and 
 a rise (and fall) time of 120 ps. In this case, the `$CV^2$' energy dissipation in the switching 
 circuit is 15 kT and the energy dissipated internally in the nanomagnet 
 due to Gilbert damping is $E_d$ = 79 kT. Finally, for cobalt,
 the minimum switching delay is 723 ps obtained with a stress of 104.5 MPa and 
 a rise (and fall) time of 50 ps. 
 The `$CV^2$' energy dissipation in the switching 
 circuit is 36 kT and the energy dissipated internally in the nanomagnet 
 due to Gilbert damping is $E_d$ = 75 kT.

All this shows that it is possible to switch multiferroic nanomagnets in less than 1 ns while dissipating 
energies of $\sim$100 kT. This range of energy dissipation is far lower than what is 
encountered in spin transfer torque based switching of nanomagnets with the same switching delay. Thus,
strain induced switching of magnets has the potential to emerge as the preferred method of 
switching magnetic bits in magnet-based non-volatile logic and memory.

%


\end{document}